\documentclass[conference, letterpaper]{IEEEtran}

\usepackage[pdftex]{graphicx}

\usepackage{xcolor}
\definecolor{burgundy}{rgb}{0.545098,0,0}
\definecolor{navyblue}{rgb}{0.0, 0.0, 0.5}
\definecolor{leafgreen}{rgb}{0.290196, 0.470588, 0.0}
\definecolor{bluegreen}{rgb}{0, 0.470588, 0.415686}
\definecolor{zuhl}{rgb}{0.1875, 0.26171875, 0.46484375}
\definecolor{orange}{rgb}{1, 0.6470588235, 0}
\definecolor{red}{rgb}{1, 0, 0}

\usepackage{ifpdf}
\usepackage[noadjust]{cite}

\usepackage{array}
\usepackage[caption=false,font=footnotesize]{subfig}
\usepackage{fixltx2e}

\usepackage[utf8]{inputenc} 
\usepackage[T1]{fontenc}
\usepackage{newtxtext}
\usepackage{url}
\usepackage{ifthen}
\usepackage{cite}
\usepackage[cmex10]{amsmath}

\usepackage{mathtools}
\usepackage{amssymb}
\usepackage{relsize}
\usepackage{bbm}
\usepackage{xfrac}
\usepackage{amsthm}
\theoremstyle{plain}
\newtheorem{definition}{Definition}
\newtheorem{lemma}{Lemma}

\newtheorem{theorem}{Theorem}
\newtheorem{corollary}{Corollary}
\newtheorem{remark}{Remark}

\newtheorem{example}{Example}

\usepackage[varg, cmbraces]{newtxmath}

\newcommand{\bvec}[1]{\boldsymbol{#1}}

\newcommand{\argmax}{\operatorname{arg~max}\limits}

\newcommand{\lemref}[1]{Lemma~\ref{#1}}

\newcommand{\thref}[1]{Theorem~\ref{#1}}
\newcommand{\defref}[1]{Definition~\ref{#1}}
\newcommand{\corref}[1]{Corollary~\ref{#1}}
\newcommand{\sectref}[1]{Section~\ref{#1}}
\newcommand{\remref}[1]{Remark~\ref{#1}}

\newcommand{\appref}[1]{Appendix~\ref{#1}}

\usepackage{overpic}
\usepackage{pict2e}
\usepackage{booktabs}

\allowdisplaybreaks[4]

\usepackage{tikz}
\usepackage{tikz-3dplot}
\usetikzlibrary{positioning}
\usetikzlibrary{decorations.pathreplacing}
\usetikzlibrary{decorations.markings}
\usetikzlibrary{positioning}
\usetikzlibrary{patterns}
\usetikzlibrary{shapes,arrows}
\tikzstyle{vecArrow} = [thick, decoration={markings,mark=at position
   1 with {\arrow[semithick]{open triangle 90}}},
   double distance=3pt, shorten >= 4pt,
   preaction = {decorate},
   postaction = {draw,line width=1.4pt, white,shorten >= 4.5pt}]
\tikzstyle{innerWhite} = [semithick, white,line width=1.4pt, shorten >= 4.5pt]
\tikzset{
    partial ellipse/.style args={#1:#2:#3}{
        insert path={+ (#1:#3) arc (#1:#2:#3)}
    }
}
\usetikzlibrary{arrows.meta}
\usetikzlibrary{fit}
\tikzset{%
  highlight/.style = {rectangle, rounded corners, fill = green!20, draw, fill opacity = 0.2, thick, inner sep = 0pt}
}

\usepackage{array}
\usepackage[linesnumbered, algoruled, boxed, lined, vlined]{algorithm2e}
\SetKwRepeat{Do}{do}{while}

\usepackage{hyperref}
\hypersetup{
    colorlinks,
    linkcolor={red!60!black},
    citecolor={blue!60!black},
    urlcolor={blue!50!black}
}
\usepackage{microtype}

\IEEEoverridecommandlockouts

\begin{document}
\flushbottom
\title{Countably Infinite Multilevel Source Polarization for Non-Stationary Erasure Distributions}

\author{%
  \IEEEauthorblockN{Yuta Sakai}
  \IEEEauthorblockA{Department of ECE,\\
                    National University of Singapore\\
                    Email: \url{eleyuta@nus.edu.sg}}
  \and
  \IEEEauthorblockN{Ken-ichi Iwata}
  \IEEEauthorblockA{Graduate School of Engineering,\\
                    University of Fukui\\
                    Email: \url{k-iwata@u-fukui.ac.jp}}
  \and
  \IEEEauthorblockN{Hiroshi Fujisaki}
  \IEEEauthorblockA{Graduate School of Natural Science and Technology,\\
                    Kanazawa University\\
                    Email: \url{fujisaki@ec.t.kanazawa-u.ac.jp}}
\thanks{This work is supported in part by JSPS KAKENHI Grant Numbers 26420352; 17K06422; 17J11247; and 18K11465, and a Singapore NRF fellowship (R-263-000-D02-281).}
}

\maketitle

\begin{abstract}
Polar transforms are central operations in the study of polar codes.
This paper examines polar transforms for non-stationary memoryless sources on \emph{possibly infinite} source alphabets.
This is the first attempt of source polarization analysis over infinite alphabets.
The source alphabet is defined to be a Polish group, and we handle the Ar{\i}kan-style two-by-two polar transform based on the group.
Defining erasure distributions based on the normal subgroup structure, we give recursive formulas of the polar transform for our proposed erasure distributions.
As a result, the recursive formulas lead to concrete examples of multilevel source polarization with \emph{countably infinite levels} when the group is locally cyclic.
We derive this result via elementary techniques in lattice theory.
\end{abstract}

\section{Introduction}

Polar codes were invented by Ar{\i}kan \cite{arikan_2009} as a provably capacity-achieving channel coding technique for binary-input memoryless symmetric channels with low complexity encoding/decoding.
The central operation in polar coding is so-called the \emph{polar transform}, which creates worse and better channels than an original channel in a certain sense.
It was shown that this coding technique can also be applied to binary source coding problems with side information \cite{arikan_isit2010}.
\c{S}a\c{s}o\u{g}lu \cite{sasoglu_2012} extended polar source coding from binary to non-binary alphabets.
While \c{S}a\c{s}o\u{g}lu's polar transform \cite[Definition~4.1]{sasoglu_2012} is non-linear in a certain sence, Mori--Tanaka \cite{mori_tanaka_2014} established non-binary polar source coding with linear polar transforms over finite fields.
In those studies, the polar transforms asymptotically create either \emph{deterministic} or \emph{equiprobable} conditional probability distributions, and these limiting proportions can be fully characterized by the conditional Shannon entropy of the original source.
Such a two-level polarization phenomenon is sometimes referred to as \emph{strong polarization.}%
\footnote{Note that the terminology ``\emph{strong polarization}'' is somewhat ambiguous in the literature. Mori--Tanaka \cite{mori_tanaka_2014} said that a channel is polarized in a weaker sense if its polar transforms do not behave as the two-level polarization; Nasser \cite{nasser_H-polarizing} said that a channel is strongly polarizing if its polar transforms behave as the two-level polarization; and B{\l}asiok et al. \cite[Definition~1.4]{blasiok_guruswami_nakkiran_rudra_sudan_2018} defined a meaning of strong polarization in a different way.}

\subsection{From Two-Level to Multilevel Polarization}

On the other hand, there are studies of investigating polarization phenomena with three or more polarization levels \cite{ergodic1, ergodic2, nasser_thesis, nasser_H-polarizing, quasigroup, fourier_analysis, sakai_iwata_itw2016, sakai_iwata_fujisaki_2018, sakai_iwata_fujisaki_isit2018, sahebi_pradhan_2013, park_barg_2013}.
Such a phenomenon is referred to as \emph{multilevel polarization.}
Roughly speaking, multilevel polarization means that the polar transform asymptotically creates equiprobable conditional distributions on some cosets of normal subgroups, provided that the polar transform is based on a finite group (cf.\ \cite[Theorem~6]{quasigroup} and \cite[Theorem~V.1]{sahebi_pradhan_2013}).
Unlike two-level polarization, characterizing the limiting proportions of multilevel polarization still remains an open problem%
\footnote{In \cite[Section~9.2.1]{nasser_thesis}, Nasser raised such an open problem aiming to find a method for calculating the exact or approximated asymptotic distribution of multilevel polarizaiton.}
in general.
In this paper, such limiting proportions are referred to as \emph{asymptotic distribution} of multilevel polarization.
In practice, the asymptotic distribution is an important indicator of constructing polar codes \cite{tal_vardy_2013}.
As a special case, the authors of this paper \cite{sakai_iwata_fujisaki_isit2018, sakai_iwata_fujisaki_2018} solved the asymptotic distribution for non-binary-input erasure-like channel models proposed in \cite{sakai_iwata_itw2016}.
Particularly, if the input alphabet size is not a prime power, then our previous works \cite{sakai_iwata_fujisaki_isit2018, sakai_iwata_fujisaki_2018} give a non-trivial method for calculating the exact asymptotic distribution algorithmically.
Recently, Nasser \cite{nasser_H-polarizing} characterized polarization levels, i.e., the support of the asymptotic distribution, for his proposed channels called \emph{automorphic-symmetric channels.}

\subsection{Binary Operation defining Polar Transform}

While the Ar{\i}kan-style two-by-two polar transform is always defined on a group of order two%
\footnote{Groups of order two are unique up to isomorphism.}
in the study of binary polar codes, note that it can be defined on several binary operations in the study of non-binary polar codes.
\c{S}a\c{s}o\u{g}lu \cite[Definition~4.1]{sasoglu_2012} considered it based on certain finite quasigroups;
Mori--Tanaka \cite{mori_tanaka_2014} on finite fields;
Park--Barg \cite{park_barg_2013} on cyclic groups in which those orders are a power of two;
Sahebi--Pradhan \cite{sahebi_pradhan_2013} on finite abelian groups;
Nasser--Telatar \cite{quasigroup} on finite quasigroups;
Nasser--Telatar \cite{fourier_analysis} also on finite abelian groups;
Nasser \cite{ergodic1, ergodic2} on more weaker finite algebras than quasigroups;
Nasser \cite{nasser_H-polarizing} also on finite, not necessarily abelian, groups;
and the authors \cite{sakai_iwata_fujisaki_isit2018, sakai_iwata_fujisaki_2018, sakai_iwata_itw2016} on finite cyclic groups.
It is worth pointing out that by the structure theorem of finite abelian groups, the polar transform based on a finite abelian group can be considered as a polar coding for the multiple-access channel (MAC) \cite{quasigroup, fourier_analysis}.
The neccesary and sufficient condition of MACs that the entire capacity region can be achieved by the MAC polarization has been characterized by Nasser--Telatar \cite{fourier_analysis} via discrete Fourier analysis over a finite abelian group.

\subsection{Related Problem: Sumset Inequality for Shannon Entropy}

Beyond the study of polar codes, bounding some measures and/or entropy of a random variable (r.v.) generated by a group action between two independent r.v.'s are central issues \cite{ruzsa_2008, tao_2010, madiman_marcus_tetali_2012, kontoyiannis_madiman_2014, guruswami_velingker_2015, abbe_li_madiman_2017, madiman_kontoyiannis_2018, li_madiman_2019}.
This problem is information-theoretically analogous to additive combinatorics or \emph{Ruzsa calculus} (cf.\ \cite{tao_vu_2006}).
Ruzsa \cite{ruzsa_2008} and Tao \cite{tao_2010} established the sumset and inverse sumset estimates on the Shannon entropy of abelian group-valued r.v.'s with countable order of the group (see also \cite{madiman_marcus_tetali_2012}).
Kontoyiannis and Madiman \cite{kontoyiannis_madiman_2014} extended such estimates to the differential entropy of real-valued r.v.'s, where the group operation is addition of real numbers.
Madiman and Kontoyiannis \cite{madiman_kontoyiannis_2018} further generalized such estimates to the differential entropy of Polish, locally compact, and abelian group-valued r.v.'s, where the probability density functions are defined with respect to a Haar measure.
Applications of such sumset inequalities for the Shannon entropy to polar codes have been discussed in the literature \cite{guruswami_velingker_2015, abbe_li_madiman_2017}.

\subsection{Contributions of This Study}

In this study, we considers the recursive application of Ar{\i}kan-style two-by-two polar transforms based on Polish groups with \emph{possibly infinite} order.
This is the first attempt of polarization analysis over infinite alphabets.
The main aim of this study is to explore the asymptotic distribution of multilevel polarization.
In the context of polar coding for a non-stationary source, we consider a mutually independent, but not necessarily identically distributed, sequence of group-valued r.v.'s with side information.
This non-stationary setting is similar to the study of binary polar coding for non-stationary channels \cite{alsan_telatar_2016, mahdavifar_2016}.
Moreover, examining the asymptotic behavior of the conditional Shannon entropy with the recursive group actions among independent r.v.'s is somewhat related to the study of sumset inequalities for the Shannon entropy \cite{ruzsa_2008, tao_2010, madiman_marcus_tetali_2012, kontoyiannis_madiman_2014, guruswami_velingker_2015, abbe_li_madiman_2017, madiman_kontoyiannis_2018, li_madiman_2019}.

We tackle this problem by proposing erasure distributions based on the normal subgroup structure of a given group, see \defref{def:group_erasure} of \sectref{sect:erasure}.
Our group-based erasure distributions are a generalization of \c{S}a\c{s}o\u{g}lu's one \cite[Section~3.3.1]{sasoglu_2012}, which is defined on the binary source alphabet.
To simplify our analysis, we derive the recursive formulas of the polar transform for the group-based erasure distributions in \thref{th:erasure_recursive} of \sectref{sect:recursive}.
When the group is locally cyclic, in \thref{th:asymptotic_distribution} of \sectref{sect:asymptotic_distribution}, we give a method of computing the exact asymptotic distribution of multilevel polarization for group-based erasure distributions.
\thref{th:asymptotic_distribution} is proved via the lattice structure of the normal subgroups \cite{ore_1938, birkhoff_1967}.
These results are more abstractly general than that of the authors' previous works \cite{sakai_iwata_itw2016, sakai_iwata_fujisaki_2018, sakai_iwata_fujisaki_isit2018}, and the first instance of \emph{countably infinite polarization levels.}
We decsribe a relation to our previous works \cite{sakai_iwata_itw2016, sakai_iwata_fujisaki_2018, sakai_iwata_fujisaki_isit2018} in \sectref{sect:MAEC};
we give the simplest case of countably infinite polarization levels with the Pr\"{u}fer $p$-group in \sectref{sect:prufer}.

\section{Problem Formulation and Basic Lemmas}

\subsection{Conditional Distribution and Conditional Shannon Entropy}

Let $(\mathcal{X}, \mathcal{B})$ be a standard Borel space, $X$ an $(\mathcal{X}, \mathcal{B})$-valued r.v., and $Y$ a r.v.
Denote by $P_{X|Y}$ a \emph{regular conditional distribution}%
\footnote{The regular conditional distribution $P_{X|Y}$ always exists, because $(\mathcal{X}, \mathcal{B})$ is standard Borel (see, e.g., \cite[Theorem~10.2.2]{dudley_2002} or \cite[Theorem~4.1.17]{durrett_2019}).}
of $X$ relative to $Y$, i.e., it is a r.v.\ forming a probability measure on $(\mathcal{X}, \mathcal{B})$ almost surely (a.s.) and $P_{X|Y}( B )$ is a version of the conditional probability $\mathbb{P}(X^{-1}(B) \mid Y)$ for each $B \in \mathcal{B}$.
\defref{def:equiv_cond} introduces an equivalence relation between two r.v.'s relative to another r.v.

\begin{definition}
\label{def:equiv_cond}
We say that two r.v.'s $Y$ and $Z$ are equivalent relative to an $(\mathcal{X}, \mathcal{B})$-valued r.v.\ $X$, denoted by $Y \equiv_{X} Z$, if
$
P_{X|Y}( B )
=
P_{X|Z}( B )
$
a.s.\ for every $B \in \mathcal{B}$.
\end{definition}

\defref{def:equiv_cond} can be reduced to the equivalence relation $\overset{\mathrm{i}}{\sim}$ introduced by Mori--Tanaka \cite[p.~2722]{mori_tanaka_2014}, provided that $\mathcal{X}$ is a finite alphabet. 
Now, we say that $X$ is \emph{conditionally discrete relative to $Y$} if there exists a $\mathcal{B}$-valued r.v.\ $D$ such that $D$ is countable a.s., and $P_{X|Y}( D ) = 1$ a.s.
Then, the conditional entropy is defined by
\begin{align}
H(X \mid Y)
\coloneqq
\mathbb{E}\bigg[ \sum_{x \in \mathcal{X}} P_{X|Y}( x ) \log \frac{ 1 }{ P_{X|Y}( x ) } \bigg] ,
\label{eq:cond_H}
\end{align}
provided that $X$ is conditionally discrete relative to $Y$.
Here, $\log$ stands for the natural logarithm satisfying $0 \log 0 = 0$.

The following lemma is trivial from the definitions.

\begin{lemma}
\label{lem:equiv}
If $Y \equiv_{X} Z$, and if $X$ is conditionally discrete relative to $Y$ or $Z$, then it holds that
\begin{align}
H(X \mid Y)
=
H(X \mid Z) .
\end{align}
\end{lemma}

Namely, the equivalence relation defined in \defref{def:equiv_cond} classifies pairs of r.v.'s having the equal conditional Shannon entropy.
\lemref{lem:equiv} can be straightforwardly extended to a more general conditional quantity \cite[Equation~(4)]{sakai_2018}.

\subsection{One-Step Polar Transform with a Polish Group}

Let $G$ be a Polish group with group operation $\bullet$, i.e., it is a topological group equipped with a complete separable metrizable topology.
Here, the group $G$ is not necessarily abelian.
Denote by $\mathcal{B}_{G}$ the Borel $\sigma$-algebra induced by the Polish topology of $G$.
Clearly, the measurable space $(G, \mathcal{B}_{G})$ is standard Borel.
Assume that the mapping $(g, h) \mapsto g \bullet h$ is Borel-measurable; and consider two independent, but not necessarily identically distributed, $(G, \mathcal{B}_{G})$-valued r.v.'s $X_{1}$ and $X_{2}$.
The one-step polar transform generates two $(G, \mathcal{B}_{G})$-valued r.v.'s $U_{1}$ and $U_{2}$ by
\begin{align}
U_{1}
& =
X_{1} \bullet X_{2} ,
\label{eq:onestep_1} \\
U_{2}
& =
X_{2} .
\label{eq:onestep_2}
\end{align}

The following lemma will be useful to simplify the subsequent analysis on the polar transform \eqref{eq:onestep_1}--\eqref{eq:onestep_2} for erasure distributions defined later in \defref{def:group_erasure}.

\begin{lemma}
\label{lem:equiv_preserved}
Given four r.v.'s $Y_{1}$, $Y_{2}$, $Z_{1}$, and $Z_{2}$, suppose that
\begin{align}
Y_{1}
& \equiv_{X_{1}}
Z_{1} ,
\\
Y_{2}
& \equiv_{X_{2}}
Z_{2} ,
\\
(X_{1}, Y_{1}, Z_{1})
& \Perp
(X_{2}, Y_{2}, Z_{2}) .
\end{align}
Then, it holds that
\begin{align}
(Y_{1}, Y_{2})
& \equiv_{U_{1}}
(Z_{1}, Z_{2}) ,
\\
(U_{1}, Y_{1}, Y_{2})
& \equiv_{U_{2}}
(U_{1}, Z_{1}, Z_{2}) .
\end{align}
\end{lemma}

\begin{IEEEproof}[Proof of \lemref{lem:equiv_preserved}]
We first prove $(U_{1}, Y_{1}, Y_{2}) \equiv_{U_{2}} (U_{1}, Z_{1}, Z_{2})$.
For each $A_{1}, A_{2} \in \mathcal{B}_{G}$, it holds that
\begin{align}
&
\mathbb{E}[ \boldsymbol{1}_{\{ U_{1} \in A_{1} \}} P_{U_{2}|U_{1}, Y_{1}, Y_{2}}( A_{2} ) \mid Y_{1}, Y_{2} ]
\notag \\
& \quad \overset{\mathclap{\text{(a)}}}{=}
\mathbb{E}[ \boldsymbol{1}_{\{ U_{1} \in A_{1} \}} \mathbb{E}[ \boldsymbol{1}_{\{ U_{2} \in A_{2} \}} \mid U_{1}, Y_{1}, Y_{2} ] \mid Y_{1}, Y_{2} ]
\notag \\
& \quad \overset{\mathclap{\text{(b)}}}{=}
\mathbb{E}[ \mathbb{E}[ \boldsymbol{1}_{\{ (U_{1}, U_{2}) \in A_{1} \times A_{2} \}} \mid U_{1}, Y_{1}, Y_{2} ] \mid Y_{1}, Y_{2} ]
\notag \\
& \quad \overset{\mathclap{\text{(c)}}}{=}
\mathbb{E}[ \boldsymbol{1}_{\{ (U_{1}, U_{2}) \in A_{1} \times A_{2} \}} \mid Y_{1}, Y_{2} ]
\notag \\
& \quad \overset{\mathclap{\text{(d)}}}{=}
\mathbb{E}[ \boldsymbol{1}_{\{ U_{2} \in A_{2} \}} P_{U_{1}|U_{2}, Y_{1}, Y_{2}}( A_{1} ) \mid Y_{1}, Y_{2} ]
\notag \\
& \quad \overset{\mathclap{\text{(e)}}}{=}
\mathbb{E}[ \boldsymbol{1}_{\{ X_{2} \in A_{2} \}} \mathbb{E}[ \boldsymbol{1}_{\{ X_{1} \bullet X_{2} \in A_{1} \}} \mid X_{2}, Y_{1}, Y_{2} ] \mid Y_{1}, Y_{2} ]
\notag \\
& \quad \overset{\mathclap{\text{(f)}}}{=}
\mathbb{E}[ \boldsymbol{1}_{\{ X_{2} \in A_{2} \}} P_{X_{1}|Y_{1}, Y_{2}}( A_{1} \bullet X_{2}^{-1} ) \mid Y_{1}, Y_{2} ]
\notag \\
& \quad \overset{\mathclap{\text{(g)}}}{=}
\mathbb{E}[ \boldsymbol{1}_{\{ X_{2} \in A_{2} \}} P_{X_{1}|Y_{1}}( A_{1} \bullet X_{2}^{-1} ) \mid Y_{1}, Y_{2} ]
\notag \\
& \quad \overset{\mathclap{\text{(h)}}}{=}
\int_{A_{2}} P_{X_{1}|Y_{1}}( A_{1} \bullet x_{2}^{-1} ) \, P_{X_{2}|Y_{1}, Y_{2}}( \mathrm{d} x_{2} )
\notag \\
& \quad \overset{\mathclap{\text{(i)}}}{=}
\int_{A_{2}} P_{X_{1}|Y_{1}}( A_{1} \bullet x_{2}^{-1} ) \, P_{X_{2}|Y_{2}}( \mathrm{d} x_{2} )
\notag \\
& \quad \overset{\mathclap{\text{(j)}}}{=}
\int_{A_{2}} P_{X_{1}|Z_{1}}( A_{1} \bullet x_{2}^{-1} ) \, P_{X_{2}|Z_{2}}( \mathrm{d} x_{2} )
\notag \\
& \quad \overset{\mathclap{\text{(k)}}}{=}
P_{U_{1}, U_{2}|Z_{1}, Z_{2}}( A_{1} \times A_{2} )
\notag \\
& \quad \overset{\mathclap{\text{(l)}}}{=}
\mathbb{E}[ \boldsymbol{1}_{\{ U_{1} \in A_{1} \}} P_{U_{2}|U_{1}, Z_{1}, Z_{2}}( A_{2} ) \mid Z_{1}, Z_{2} ]
\label{eq:equiv_U2}
\end{align}
a.s., where
(a) follows by the definition of conditional probabilities;
(b) follows from the fact that $\boldsymbol{1}_{\{ U_{1} \in A_{1} \}}$ is $\sigma(U_{1}, Y_{1}, Y_{2})$-measurable;
(c) follows from the fact that
\begin{align}
\mathcal{G} \subset \mathcal{H}
\ \Longrightarrow \
\mathbb{E}[ Z \mid \mathcal{G} ] = \mathbb{E}[ \mathbb{E}[ Z \mid \mathcal{H} ] \mid \mathcal{G} ]
\quad \text{(a.s.)}
\label{eq:cond_nesting}
\end{align}
for a real-valued r.v.\ $Z$ (cf.\ \cite[Theorem~9.1.5]{chung_2001});
(d) follows as in (a)--(c);
(e) follows by \eqref{eq:onestep_1}--\eqref{eq:onestep_2};
(f) follows from $X_{1} \Perp X_{2} \mid (Y_{1}, Y_{2})$ together with \cite[Theorem~9.2.2]{chung_2001};
(g) follows from \cite[Theorem~9.2.1]{chung_2001} and the fact that $(X_{1}, Y_{1}) \Perp (X_{2}, Y_{2})$ implies $X_{1} \Perp Y_{2} \mid Y_{1}$;
(h) follows since $P_{X_{2}|Y_{1}, Y_{2}}( \cdot )$ forms a probability measure on $(G, \mathcal{B}_{G})$ a.s.;
(i) follows from \cite[Theorem~9.2.1]{chung_2001} and the fact that $(X_{1}, Y_{1}) \Perp (X_{2}, Y_{2})$ implies $X_{2} \Perp Y_{1} \mid Y_{2}$;
(j) follows by $Y_{1} \equiv_{X_{1}} Z_{1}$ and $Y_{2} \equiv_{X_{2}} Z_{2}$;
(k) follows as in (d)--(f); and
(l) follows as in (a)--(c).
Namely, Equation~\eqref{eq:equiv_U2} is equivalent to
$
P_{U_{2}|U_{1}, Y_{1}, Y_{2}}( A_{2} )
=
P_{U_{2}|U_{1}, Z_{1}, Z_{2}}( A_{2} )
$
a.s.\ for $A_{2} \in \mathcal{B}_{G}$;
and thus, we have $(U_{1}, Y_{1}, Y_{2}) \equiv_{U_{2}} (U_{1}, Z_{1}, Z_{2})$, as desired.

We next prove $(Y_{1}, Y_{2}) \equiv_{U_{1}} (Z_{1}, Z_{2})$.
By setting $A_{2} = G$, it can be verified from \eqref{eq:equiv_U2} that for each $A_{1} \in \mathcal{B}_{G}$,
\begin{align}
P_{U_{1} | Y_{1}, Y_{2}}( A_{1} )
& =
\int_{G} P_{X_{1}|Y_{1}}( A_{1} \bullet x_{2}^{-1} ) \, P_{X_{2}|Y_{2}}( \mathrm{d} x_{2} )
\notag \\
& =
\int_{G} P_{X_{1}|Z_{1}}( A_{1} \bullet x_{2}^{-1} ) \, P_{X_{2}|Z_{2}}( \mathrm{d} x_{2} )
\notag \\
& =
P_{U_{1} | Z_{1}, Z_{2}}( A_{1} )
\label{eq:equiv_U1}
\end{align}
a.s.
Thus, we have $(Y_{1}, Y_{2}) \equiv_{U_{1}} (Z_{1}, Z_{2})$, as desired.
\end{IEEEproof}

\lemref{lem:equiv_preserved} means that the polar transform \eqref{eq:onestep_1}--\eqref{eq:onestep_2} preserves the equivalence relation defined in \defref{def:equiv_cond}.
In the context of polar channel codes, as an analoguos result to \lemref{lem:equiv_preserved}, the preserving property of a certain channel ordering was firstly shown by Korada \cite[Lemma~4.7]{korada_thesis} (see also \cite[Lemma~5]{tal_vardy_2013}), and its generalization from binary to arbitrary input alphabets was given in \cite[Lemma~2]{sakai_iwata_fujisaki_2018} under a quasigroup operation.

\subsection{Recursive Construction of Polar Transforms}
\label{sect:recursive}

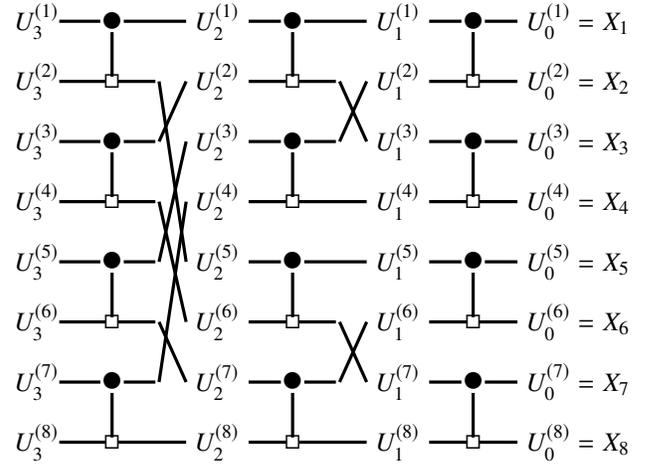
\begin{figure}[!t]
\centering
\begin{tikzpicture}[scale = 0.8]
\node (x1) at (0, 0) {$U_{0}^{(1)} = X_{1}$};
\node (x2) at (0, -1) {$U_{0}^{(2)} = X_{2}$};
\node (x3) at (0, -2) {$U_{0}^{(3)} = X_{3}$};
\node (x4) at (0, -3) {$U_{0}^{(4)} = X_{4}$};
\node (x5) at (0, -4) {$U_{0}^{(5)} = X_{5}$};
\node (x6) at (0, -5) {$U_{0}^{(6)} = X_{6}$};
\node (x7) at (0, -6) {$U_{0}^{(7)} = X_{7}$};
\node (x8) at (0, -7) {$U_{0}^{(8)} = X_{8}$};
\node (plus11) at (-1.75, 0) [inner sep = 0pt] {\LARGE $\bullet$};
\node (plus12) at (-1.75, -2) [inner sep = 0pt] {\LARGE $\bullet$};
\node (plus13) at (-1.75, -4) [inner sep = 0pt] {\LARGE $\bullet$};
\node (plus14) at (-1.75, -6) [inner sep = 0pt] {\LARGE $\bullet$};
\node (dot11) at (-1.75, -1) [inner sep = -1pt] {$\square$};
\node (dot12) at (-1.75, -3) [inner sep = -1pt] {$\square$};
\node (dot13) at (-1.75, -5) [inner sep = -1pt] {$\square$};
\node (dot14) at (-1.75, -7) [inner sep = -1pt] {$\square$};
\node (u11) at (-3, 0) {$U_{1}^{(1)}$};
\node (u12) at (-3, -1) {$U_{1}^{(2)}$};
\node (u13) at (-3, -2) {$U_{1}^{(3)}$};
\node (u14) at (-3, -3) {$U_{1}^{(4)}$};
\node (u15) at (-3, -4) {$U_{1}^{(5)}$};
\node (u16) at (-3, -5) {$U_{1}^{(6)}$};
\node (u17) at (-3, -6) {$U_{1}^{(7)}$};
\node (u18) at (-3, -7) {$U_{1}^{(8)}$};
\draw [very thick] (x1) -- (plus11) -- (u11);
\draw [very thick] (x3) -- (plus12) -- (u13);
\draw [very thick] (x5) -- (plus13) -- (u15);
\draw [very thick] (x7) -- (plus14) -- (u17);
\draw [very thick] (x2) -- (dot11) -- (u12);
\draw [very thick] (x4) -- (dot12) -- (u14);
\draw [very thick] (x6) -- (dot13) -- (u16);
\draw [very thick] (x8) -- (dot14) -- (u18);
\draw [very thick] (plus11) -- (dot11);
\draw [very thick] (plus12) -- (dot12);
\draw [very thick] (plus13) -- (dot13);
\draw [very thick] (plus14) -- (dot14);
\node (v12) at (-4, -1) [inner sep = 0pt] {};
\node (v13) at (-4, -2) [inner sep = 0pt] {};
\node (v16) at (-4, -5) [inner sep = 0pt] {};
\node (v17) at (-4, -6) [inner sep = 0pt] {};
\draw [very thick] (u13.west) -- (v12.east);
\draw [very thick] (u12.west) -- (v13.east);
\draw [very thick] (u17.west) -- (v16.east);
\draw [very thick] (u16.west) -- (v17.east);
\node (plus21) at (-4.75, 0) [inner sep = 0pt] {\LARGE $\bullet$};
\node (plus22) at (-4.75, -2) [inner sep = 0pt] {\LARGE $\bullet$};
\node (plus23) at (-4.75, -4) [inner sep = 0pt] {\LARGE $\bullet$};
\node (plus24) at (-4.75, -6) [inner sep = 0pt] {\LARGE $\bullet$};
\node (dot21) at (-4.75, -1) [inner sep = -1pt] {$\square$};
\node (dot22) at (-4.75, -3) [inner sep = -1pt] {$\square$};
\node (dot23) at (-4.75, -5) [inner sep = -1pt] {$\square$};
\node (dot24) at (-4.75, -7) [inner sep = -1pt] {$\square$};
\node (u21) at (-6, 0) {$U_{2}^{(1)}$};
\node (u22) at (-6, -1) {$U_{2}^{(2)}$};
\node (u23) at (-6, -2) {$U_{2}^{(3)}$};
\node (u24) at (-6, -3) {$U_{2}^{(4)}$};
\node (u25) at (-6, -4) {$U_{2}^{(5)}$};
\node (u26) at (-6, -5) {$U_{2}^{(6)}$};
\node (u27) at (-6, -6) {$U_{2}^{(7)}$};
\node (u28) at (-6, -7) {$U_{2}^{(8)}$};
\draw [very thick] (u11) -- (plus21) -- (u21);
\draw [very thick] (v13) -- (plus22) -- (u23);
\draw [very thick] (u15) -- (plus23) -- (u25);
\draw [very thick] (v17) -- (plus24) -- (u27);
\draw [very thick] (v12) -- (dot21) -- (u22);
\draw [very thick] (u14) -- (dot22) -- (u24);
\draw [very thick] (v16) -- (dot23) -- (u26);
\draw [very thick] (u18) -- (dot24) -- (u28);
\draw [very thick] (plus21) -- (dot21);
\draw [very thick] (plus22) -- (dot22);
\draw [very thick] (plus23) -- (dot23);
\draw [very thick] (plus24) -- (dot24);
\node (v22) at (-7, -1) [inner sep = 0pt] {};
\node (v23) at (-7, -2) [inner sep = 0pt] {};
\node (v24) at (-7, -3) [inner sep = 0pt] {};
\node (v25) at (-7, -4) [inner sep = 0pt] {};
\node (v26) at (-7, -5) [inner sep = 0pt] {};
\node (v27) at (-7, -6) [inner sep = 0pt] {};
\draw [very thick] (u25.west) -- (v22.east);
\draw [very thick] (u22.west) -- (v23.east);
\draw [very thick] (u26.west) -- (v24.east);
\draw [very thick] (u23.west) -- (v25.east);
\draw [very thick] (u27.west) -- (v26.east);
\draw [very thick] (u24.west) -- (v27.east);
\node (plus31) at (-7.75, 0) [inner sep = 0pt] {\LARGE $\bullet$};
\node (plus32) at (-7.75, -2) [inner sep = 0pt] {\LARGE $\bullet$};
\node (plus33) at (-7.75, -4) [inner sep = 0pt] {\LARGE $\bullet$};
\node (plus34) at (-7.75, -6) [inner sep = 0pt] {\LARGE $\bullet$};
\node (dot31) at (-7.75, -1) [inner sep = -1pt] {$\square$};
\node (dot32) at (-7.75, -3) [inner sep = -1pt] {$\square$};
\node (dot33) at (-7.75, -5) [inner sep = -1pt] {$\square$};
\node (dot34) at (-7.75, -7) [inner sep = -1pt] {$\square$};
\node (u31) at (-9, 0) [inner sep = 0pt] {$U_{3}^{(1)}$};
\node (u32) at (-9, -1) [inner sep = 0pt] {$U_{3}^{(2)}$};
\node (u33) at (-9, -2) [inner sep = 0pt] {$U_{3}^{(3)}$};
\node (u34) at (-9, -3) [inner sep = 0pt] {$U_{3}^{(4)}$};
\node (u35) at (-9, -4) [inner sep = 0pt] {$U_{3}^{(5)}$};
\node (u36) at (-9, -5) [inner sep = 0pt] {$U_{3}^{(6)}$};
\node (u37) at (-9, -6) [inner sep = 0pt] {$U_{3}^{(7)}$};
\node (u38) at (-9, -7) [inner sep = 0pt] {$U_{3}^{(8)}$};
\draw [very thick] (u21) -- (plus31) -- (u31);
\draw [very thick] (v23) -- (plus32) -- (u33);
\draw [very thick] (v25) -- (plus33) -- (u35);
\draw [very thick] (v27) -- (plus34) -- (u37);
\draw [very thick] (v22) -- (dot31) -- (u32);
\draw [very thick] (v24) -- (dot32) -- (u34);
\draw [very thick] (v26) -- (dot33) -- (u36);
\draw [very thick] (u28) -- (dot34) -- (u38);
\draw [very thick] (plus31) -- (dot31);
\draw [very thick] (plus32) -- (dot32);
\draw [very thick] (plus33) -- (dot33);
\draw [very thick] (plus34) -- (dot34);
\end{tikzpicture}
\caption{Recursive construction of polar transforms up to $n \le 3$ (see \eqref{eq:polar_transform_minus}--\eqref{eq:polar_transform_plus}).}
\label{fig:recursive_construction}
\end{figure}

This subsection introduces the recursive applications of the polar transform \eqref{eq:onestep_1}--\eqref{eq:onestep_2}.
For a sequence $\{ X_{i} \}_{i = 1}^{\infty}$ of $(G, \mathcal{B}_{G})$-valued r.v.'s, we recursively construct the double sequence $\{ U_{n}^{(i)} \}_{i = 1 ,}^{\infty} {}_{n = 0}^{\infty}$ of r.v.'s by%
\footnote{The recursive formulas \eqref{eq:polar_transform_minus}--\eqref{eq:polar_transform_plus} for binary-input non-starionary channels can be found in, e.g.,  \cite[Section~III]{alsan_telatar_2016} and \cite[Equation~(7)]{mahdavifar_2016}.}
\begin{align}
U_{n}^{(2 i - 1)}
& \coloneqq
U_{n-1}^{(i)} \bullet U_{n-1}^{(i + 2^{n-1})} ,
\label{eq:polar_transform_minus} \\
U_{n}^{(2 i)}
& \coloneqq
U_{n-1}^{(i + 2^{n-1})}
\label{eq:polar_transform_plus}
\end{align}
for each $i \ge 1$ and $n \ge 1$, where $\{ U_{0}^{(i)} \}_{i = 1}^{\infty} \coloneqq \{ X_{i} \}_{i = 1}^{\infty}$.
Figure~\ref{fig:recursive_construction} illustrates a diagram of \eqref{eq:polar_transform_minus} and \eqref{eq:polar_transform_plus}.
Note that $U_{n}^{(i)}$ is also $(G, \mathcal{B}_{G})$-valued for all $n \ge 0$ and $i \ge 1$.

As a non-stationary source with side information, consider a sequence $\{ Y_{i} \}_{i = 1}^{\infty}$ of r.v.'s playing the role of side information for $\{ X_{i} \}_{i = 1}^{\infty}$.
Suppose that the sequence $\{ (X_{i}, Y_{i}) \}_{i = 1}^{\infty}$ is mutually independent.
Define the double sequence $\{ \mathfrak{H}_{n}^{(i)} \}_{i = 1 ,}^{\infty} {}_{n = 0}^{\infty}$ of conditional Shannon entropies by
\begin{align}
\mathfrak{H}_{n}^{(i)}
\coloneqq
H(U_{n}^{(i)} \mid \{ U_{n}^{(j)} \}_{j = 1}^{i-1}, \{ Y_{j} \}_{j = 1}^{\infty})
\label{def:mathfrak_H}
\end{align}
for each $n \ge 0$ and $i \ge 1$.
Roughly speaking, a source polarization theorem explores how the sequence $\{ \mathfrak{H}_{n}^{(i)} \}_{i = 1}^{\infty}$ behaves for sufficiently large $n$.
It can be verified by \eqref{eq:polar_transform_minus}--\eqref{eq:polar_transform_plus} that
$U_{n}^{(i)}$
is conditionally independent of
$(\{ U_{n}^{(j)} \}_{j = 1}^{m 2^{n}},  \{ Y_{j} \}_{j = 1}^{m 2^{n}}, \{ Y_{j} \}_{j = (m+1) 2^{n}+1}^{\infty})$
relative to
$(\{ U_{n}^{(j)} \}_{j = m 2^{n}+1}^{i}, \{ Y_{j} \}_{j = m 2^{n}+1}^{(m+1)2^{n}})$,
where $m \coloneqq \lfloor i / 2^{n} \rfloor$, and $\lfloor \cdot \rfloor$ stands for the floor function.
Namely, it holds that
\begin{align}
\mathfrak{H}_{n}^{(i)}
=
H( U_{n}^{(i)} \mid \{ U_{n}^{(j)} \}_{j = m 2^{n}+1}^{i-1}, \{ Y_{j} \}_{j = m 2^{n}+1}^{(m+1)2^{n}} ) .
\label{eq:dependence}
\end{align}
The next section analyzes the conditional Shannon entropies $\{ \mathfrak{H}_{n}^{(i)} \}_{i = 1 ,}^{\infty} {}_{n = 0}^{\infty}$ for erasure distributions.

\section{Polar Transforms for Erasure Distributions}

\subsection{Erasure Distribution Based on Group Structure}
\label{sect:erasure}

It is well-known that the polar transform for binary erasure channels (BECs) can be easily analyzed \cite[Proposition~6]{arikan_2009} by a certain recursive formula.
This convenient probabilistic model was translated from channel to source coding problems by \c{S}a\c{s}o\u{g}lu \cite[Section~3.3.1]{sasoglu_2012} by defining an \emph{erasure distribution} as an analogy to the BEC.
In this subsection, from the group theoretic perspective, we introduce a more general erasure distribution than \c{S}a\c{s}o\u{g}lu's one.

Let $H \lhd G$ be a shorthand for a normal subgroup $H$ of $G$.
The following definition gives an erasure distribution based on the normal subgroup structure of $G$.

\begin{definition}[group-based erasure distribution]
\label{def:group_erasure}
Let $X$ be an $(G, \mathcal{B}_{G})$-valued r.v., and $Y$ a $\mathcal{Y}$-valued r.v., where
\begin{align}
\mathcal{Y}
& =
\bigcup_{ H \lhd G: \text{$H$ $\mathrm{is}$ $\mathrm{finite}$} } \frac{ G }{ H }
=
\{ g \bullet H \mid g \in G \ \mathrm{and} \ \mathrm{finite} \ H \lhd G \} .
\notag
\end{align}
Then, we say that $(X, Y)$ follows an \emph{erasure distribution} if
\begin{align}
P_{X|Y}( B )
=
\frac{ |B \cap Y| }{ |Y| }
\quad \mathrm{(a.s.)}, \qquad \mathrm{for} \ B \in \mathcal{B}_{G} .
\label{eq:coset_erasure}
\end{align}
\end{definition}

The alphabet $\mathcal{Y}$ given in \defref{def:group_erasure} is the collection of finite cosets of $G$, and note that $X \in Y$ a.s.\ if $(X, Y)$ follows an erasure distribution.
Intuitively, this means that the side information $Y$ only tells us that $X$ belongs to a finite coset $Y$.
It is clear that $X$ is conditionally discrete relative to $Y$ if $(X, Y)$ follows an erasure distribution, because $P_{X|Y}$ is a uniform distribution on a finite coset $Y$ a.s.\ (see \eqref{eq:coset_erasure}).

If the order of $G$ is two, then \defref{def:group_erasure} coincides with \c{S}a\c{s}o\u{g}lu's one \cite[Section~3.3.1]{sasoglu_2012}.
Moreover, \defref{def:group_erasure} can be reduced to some know erasure-like channels as follows.

\begin{remark}
\label{rem:reduction}
Let $(X, Y)$ a pair of r.v.'s following an erasure distribution in the sense of \defref{def:group_erasure}.
Suppose that
\begin{align}
\mathbb{P}\{ Y = g \bullet H \mid Y \in G/H \}
=
\mathbb{P}\{ Y = H \mid Y \in G/H \}
\end{align}
for every $g \in G$ and every finite $H \lhd G$, provided that $\mathbb{P}\{ Y \in G/H \} > 0$.
If $G$ is a finite cyclic group, then the joint probability measure induced by $(X, Y)$ is equivalent to a \emph{modular arithmetic erasure channel} \cite[Definition~2]{sakai_iwata_fujisaki_2018} with uniform input distribution.
As summarized in \cite[Examples~1--4]{sakai_iwata_fujisaki_2018}, modular arithmetic erasure channels can be further reduced to many other erasure-like channels given in \cite{park_barg_isit2011, sahebi_pradhan_2013}.
In addition, if $G$ is a finite elementary abelian group, then the joint probability measure induced by $(X, Y)$ is equivalent to a \emph{combination of linear channels} \cite[Definition~23]{sakai_iwata_fujisaki_2018} with uniform input distribution, where the linear channels are constructed on the set of subspaces of a finite-dimensional vector space over the prime field.
\end{remark}

\lemref{lem:Q_Y} shows a simple formula for $H(X \mid Y)$ for an erasure distribution; it can be directly proven by \eqref{eq:cond_H} and \defref{def:group_erasure}.

\begin{lemma}
\label{lem:Q_Y}
If $(X, Y)$ follows an erasure distribution, then
\begin{align}
H(X \mid Y)
=
\mathbb{E}[ \log |Y| ] .
\end{align}
\end{lemma}

It is worth pointing out that \lemref{lem:Q_Y} is analogous to \cite[Proposition~3]{quasigroup} and \cite[Proposition~2]{sakai_iwata_fujisaki_2018}.

\subsection{Reduction of Polar Transforms to Erasure Distributions}
\label{sect:recursive}

Suppose that $(X_{i}, Y_{i})$ follows an erasure distribution for $i \ge 1$.
Let $\{ Y_{n}^{(i)} \}_{i = 1 ,}^{\infty} {}_{n = 0}^{\infty}$ be a double sequence of r.v.'s defined by%
\footnote{Assume w.l.o.g.\ that $Y_{n}^{(2 i)} \coloneqq \{ e \}$ if $\phi(U_{n}^{(2n - 1)}, Y_{n-1}^{(i)} , Y_{n-1}^{(i+2^{n-1})}) = \emptyset$.}
\begin{align}
Y_{n}^{(2 i - 1)}
& \coloneqq
Y_{n-1}^{(i)} \bullet Y_{n-1}^{(i+2^{n-1})} ,
\\
Y_{n}^{(2 i)}
& \coloneqq
\phi(U_{n}^{(2n - 1)}, Y_{n-1}^{(i)} , Y_{n-1}^{(i+2^{n-1})})
\label{def:Y_plus}
\end{align}
for each $i \ge 1$ and $n \ge 1$, where $\{ Y_{0}^{(i)} \}_{i = 1}^{\infty} \coloneqq \{ Y_{i} \}_{i = 1}^{\infty}$, and the mapping $\phi : G \times \mathcal{Y} \times \mathcal{Y} \to \mathcal{Y} \cup \{ \emptyset \}$ is defined by
\begin{align}
\phi : (g, a \bullet H, b \bullet K) \mapsto (a^{-1} \bullet g \bullet H) \cap (b \bullet K)
\end{align}
for $a, b, g \in G$ and finite $H, K \lhd G$.
\thref{th:erasure_recursive} shows that like the BEC \cite[Proposition~6]{arikan_2009}, the polar transform for erasure distributions generates other erasure distributions again.

\begin{theorem}
\label{th:erasure_recursive}
The pair $(U_{n}^{(i)}, Y_{n}^{(i)})$ also follows an erasure distribution for each $i \ge 1$ and $n \ge 0$.
In addition, for each $n \ge 0$ and $i \ge 1$, and with $m \coloneqq \lfloor i / 2^{n} \rfloor$, it holds that
\begin{align}
\!
Y_{n}^{(i)}
\equiv_{U_{n}^{(i)}}
(\{ U_{n}^{(j)} \}_{j = m 2^{n}+1}^{i-1}, \{ Y_{j} \}_{j = m 2^{n}+1}^{(m+1)2^{n}}) .
\end{align}
\end{theorem}

\begin{corollary}
\label{cor:erasure_recursive}
If $(X_{i}, Y_{i})$ follows an erasure distribution, then
\begin{align}
\mathfrak{H}_{n}^{(i)}
=
\mathbb{E}[ \log |Y_{n}^{(i)}| ]
\qquad \mathrm{for} \ n \ge 0, \ i \ge 1 .
\end{align}
\end{corollary}

\begin{IEEEproof}[Proof of \thref{th:erasure_recursive}]
By \lemref{lem:equiv_preserved} and the recursivity of \eqref{eq:polar_transform_minus}--\eqref{eq:polar_transform_plus}, it suffices to consider the one-step polar transform \eqref{eq:onestep_1}--\eqref{eq:onestep_2}.
Next, we prove \thref{th:erasure_recursive} in two parts:

\subsubsection{Proof for Minus Transform $(U_{1}, (Y_{1}, Y_{2}))$}

Since $X_{1}$ and $X_{2}$ are conditionally discrete relative to $Y_{1}$ and $Y_{2}$, respectively, it follows by the first equality of \eqref{eq:equiv_U1} that
\begin{align}
\!\!
P_{U_{1}|Y_{1}, Y_{2}}( A_{1} )
& =
\sum_{u_{2} \in G} P_{X_{1} | Y_{1}}( A_{1} \bullet u_{2}^{-1} ) \, P_{X_{2} | Y_{2}}( u_{2} )
\notag \\
& =
\sum_{u_{1} \in A_{1}} \sum_{u_{2} \in G} P_{X_{1} | Y_{1}}( u_{1} \bullet u_{2}^{-1} ) \, P_{X_{2} | Y_{2}}( u_{2} )
\label{eq:P_U1|Y1Y2_single}
\end{align}
a.s.\ for every $A_{1} \in \mathcal{B}_{G}$.
Thus, we see that $U_{1}$ is conditionally discrete relative to $(Y_{1}, Y_{2})$ as well.
It follows from \eqref{eq:coset_erasure} that
\begin{align}
P_{X_{1}|Y_{1}}( u_{1} \bullet u_{2}^{-1} ) \, P_{X_{2}|Y_{2}}( u_{2} )
& =
\frac{ \boldsymbol{1}_{\{ u_{1} \bullet u_{2}^{-1} \in Y_{1} \} \cap \{ u_{2} \in Y_{2} \}} }{ |Y_{1}| |Y_{2}| }
\label{eq:product_regular_erasure}
\end{align}
a.s.\ for every $(u_{1}, u_{2}) \in G^{2}$.
Now, we readily see that for some $u_{1}, u_{2}, g_{1}, g_{2} \in G$ and $H, K \lhd G$, both $u_{1} \bullet u_{2}^{-1} \in g_{1} \bullet H$ and $u_{2} \in g_{2} \bullet K$ hold if and only if the following system of two congruences holds:
\begin{align}
\left\{
\begin{array}{ll}
u_{2}
\equiv
g_{1}^{-1} \bullet u_{1}
& \pmod{H} ,
\\
u_{2}
\equiv
g_{2}
& \pmod{K} .
\end{array}
\right.
\label{equiv:u2}
\end{align}
By the Chinese Remainder Theorem in group theory, the system \eqref{equiv:u2} has a unique solution $u_{2} \in G$ modulo $H \cap K$ if and only if
$
u_{1}
\equiv
g_{1} \bullet g_{2} \pmod{H \bullet K}
$.
Therefore, we have
\begin{align}
\{ u_{2} \in G \mid (u_{1} \bullet u_{2}^{-1}, u_{2}) \in Y_{1} \times Y_{2}\}
\in
\frac{ G }{ H \cap K }
\label{set_relation:CRT}
\end{align}
if and only if $Y_{1} \in G/H$, $Y_{2} \in G/K$, and $u_{1} \in Y_{1} \bullet Y_{2}$.
Hence, it follows from \eqref{eq:P_U1|Y1Y2_single}, \eqref{eq:product_regular_erasure}, and \eqref{set_relation:CRT} that for each $A_{1} \in \mathcal{B}_{G}$,
\begin{align}
\!\!
P_{U_{1}|Y_{1}, Y_{2}}( A_{1} )
& =
\sum_{u_{1} \in A_{1} \cap (Y_{1} \bullet Y_{2})} \sum_{u_{2} \in \phi(U_{1}, Y_{1}, Y_{2})} \frac{ 1 }{ |Y_{1}| \, |Y_{2}| }
\notag \\
& =
\sum_{u_{1} \in A_{1} \cap (Y_{1} \bullet Y_{2})} \frac{ |\phi(U_{1}, Y_{1}, Y_{2})| }{ |Y_{1}| \, |Y_{2}| }
\notag \\
& =
\frac{ |A_{1} \cap (Y_{1} \bullet Y_{2})| }{ |Y_{1} \bullet Y_{2}| }
\label{P_U1|Y1Y2_final}
\end{align}
a.s., where the last equality follows by the identity%
\footnote{Consider the second isomorphism theorem \emph{without} the topological structure.}
\begin{align}
|H \bullet K| \, |H \cap K|
=
|H| \, |K|
\label{eq:2nd-isomorphism}
\end{align}
for finite $H, K \lhd G$.
Equation~\eqref{P_U1|Y1Y2_final} implies that $(U_{1}, (Y_{1}, Y_{2}))$ follows an erasure distribution (see \eqref{eq:coset_erasure}).
Furthermore, since $Y_{1} \bullet Y_{2}$ is a function of $(Y_{1}, Y_{2})$, we see that $\sigma(Y_{1} \bullet Y_{2}) \subset \sigma(Y_{1}, Y_{2})$.
Hence, it can be verified by \eqref{eq:cond_nesting} and \eqref{P_U1|Y1Y2_final} that
\begin{align}
Y_{1} \bullet Y_{2}
\equiv_{U_{1}}
(Y_{1}, Y_{2}) ,
\end{align}
completing the proof for minus transform.

\subsubsection{Proof for Plus Transform $(U_{2}, (U_{1}, Y_{1}, Y_{2}))$}

Since $X_{1}$ and $X_{2}$ are conditionally discrete relative to $Y_{1}$ and $Y_{2}$, respectively, it follows as in Steps~(a)--(i) of \eqref{eq:equiv_U2} that for each $A_{1}, A_{2} \in \mathcal{B}_{G}$,
\begin{align}
&
\mathbb{E}[ \boldsymbol{1}_{\{ U_{1} \in A_{1} \}} P_{U_{2}|U_{1}, Y_{1}, Y_{2}}( A_{2} ) \mid Y_{1}, Y_{2} ]
\notag \\
& \qquad =
\sum_{u_{1} \in A_{1}} \sum_{u_{2} \in A_{2}} P_{X_{1}|Y_{1}}( u_{1} \bullet u_{2}^{-1} ) \, P_{X_{2}|Y_{1}}( u_{2} )
\notag \\
& \qquad \overset{\mathclap{\text{(a)}}}{=}
\sum_{u_{1} \in A_{1} \cap (Y_{1} \bullet Y_{2})} \sum_{u_{2} \in A_{2} \cap \phi(U_{1}, Y_{1}, Y_{2})} \frac{ 1 }{ |Y_{1}| \, |Y_{2}| }
\notag \\
& \qquad =
\sum_{u_{1} \in A_{1} \cap (Y_{1} \bullet Y_{2})} \frac{ |A_{2} \cap \phi(U_{1}, Y_{1}, Y_{2})| }{ |Y_{1}| \, |Y_{2}| }
\notag \\
& \qquad \overset{\mathclap{\text{(b)}}}{=}
\sum_{u_{1} \in A_{1} \cap (Y_{1} \bullet Y_{2})} \frac{ P_{U_{1}|Y_{1}, Y_{2}}( u_{1} ) }{ P_{U_{1}|Y_{1}, Y_{2}}( u_{1} ) } \frac{ |A_{2} \cap \phi(U_{1}, Y_{1}, Y_{2})| }{ |Y_{1}| \, |Y_{2}| }
\notag \\
& \qquad \overset{\mathclap{\text{(c)}}}{=}
\sum_{u_{1} \in A_{1} \cap (Y_{1} \bullet Y_{2})} P_{U_{1}|Y_{1}, Y_{2}}( u_{1} ) \, \frac{ |A_{2} \cap \phi(U_{1}, Y_{1}, Y_{2})| }{ |\phi(U_{1}, Y_{1}, Y_{2})| }
\notag \\
& \qquad \overset{\mathclap{\text{(d)}}}{=}
\mathbb{E}\bigg[ \boldsymbol{1}_{\{ U_{1} \in A_{1} \}} \frac{ |A_{2} \cap \phi(U_{1}, Y_{1}, Y_{2})| }{ |\phi(U_{1}, Y_{1}, Y_{2})| } \ \bigg| \ Y_{1}, Y_{2} \bigg]
\end{align}
a.s., where
(a) follows from \eqref{eq:product_regular_erasure} and \eqref{set_relation:CRT};
(b) follows from the fact that $P_{U_{1}|Y_{1}, Y_{2}}( u_{1} ) > 0$ if and only if $u_{1} \in Y_{1} \bullet Y_{2}$ (see \eqref{P_U1|Y1Y2_final});
(c) follows from \eqref{P_U1|Y1Y2_final}--\eqref{eq:2nd-isomorphism}; and
(d) follows from the fact that $P_{U_{1}|Y_{1}, Y_{2}}( \cdot )$ forms a uniform distribution on $Y_{1} \bullet Y_{2}$ a.s.
Therefore, it holds that for each $A_{2} \in \mathcal{B}_{G}$,
\begin{align}
P_{U_{2}|U_{1}, Y_{1}, Y_{2}}( A_{2} )
=
\frac{ |A_{2} \cap \phi(U_{1}, Y_{1}, Y_{2})| }{ |\phi(U_{1}, Y_{1}, Y_{2})| }
\label{P_U2|U1Y1Y2_final}
\end{align}
a.s.
Equation~\eqref{P_U2|U1Y1Y2_final} implies that $(U_{2}, (U_{1}, Y_{1}, Y_{2}))$ follows an erasure distribution.
Furthermore, since $\phi(U_{1}, Y_{1}, Y_{2})$ is a function of $(U_{1}, Y_{1}, Y_{2})$, it holds that $\sigma( \phi(U_{1}, Y_{1}, Y_{2}) ) \subset \sigma(U_{1}, Y_{1}, Y_{2})$.
Hence, it can be verified by \eqref{eq:cond_nesting} and \eqref{P_U2|U1Y1Y2_final} that
\begin{align}
\phi(U_{1}, Y_{1}, Y_{2})
\equiv_{U_{2}}
(U_{1}, Y_{1}, Y_{2}) ,
\end{align}
completing the proof for plus transform.
\end{IEEEproof}

\begin{IEEEproof}[Proof of \corref{cor:erasure_recursive}]
\corref{cor:erasure_recursive} is now obvious from Lemmas~\ref{lem:equiv}--\ref{lem:Q_Y} and \thref{cor:erasure_recursive}.
\end{IEEEproof}

If the order of $G$ is two, then \thref{th:erasure_recursive} can be reduced to the discussion in \cite[Section~3.3.1]{sasoglu_2012}, which is alanogous to the ease of analysing the polar transform for BECs \cite[Proposition~6]{arikan_2009}.
As shown in the following remark, \thref{th:erasure_recursive} is indeed a generalization of the known formulas in easy case studies of Ar{\i}kan-style two-by-two polar transforms.

\begin{remark}
\label{rem:reduction2}
For a source $\{ (X_{i}, Y_{i}) \}_{i = 1}^{\infty}$, suppose the same hypothesis as \remref{rem:reduction}.
If $G$ is a finite cyclic group, then \thref{th:erasure_recursive} is a counterpart of the recursive formula for modular arithmetic erasure channels \cite[Theorem~1]{sakai_iwata_fujisaki_2018}.
In addition, if $G$ is a finite elementary abelian group, then \thref{th:erasure_recursive} is a counterpart of the recursive formula for combinations of linear channels \cite[Proposition~4]{quasigroup} in a stationary setting.
\end{remark}

By \corref{cor:erasure_recursive}, the probability measures induced by $\{ |Y_{n}^{(i)}| \}_{i = 1}^{\infty}$ are important to analyze the asymptotic distribution of multilevel polarization for sufficiently large $n$.
The next section explores the asymptotic distribution in a special case.

\section{Multilevel Source Polarization Analysis}

A group $G$ is said to be \emph{locally cyclic} if every finitely generated subgroup of $G$ is cyclic.
In this section, we investigate the multilevel polarization theorem for a non-stationary source $\{ (X_{i}, Y_{i}) \}_{i = 1}^{\infty}$, where $(X_{i}, Y_{i})$ follows an erasure distribution with a locally cyclic group $G$ for each $i \ge 1$.

\subsection{\thref{th:erasure_recursive} with a Locally Cyclic Group}

It is known that every locally cyclic group is isomorphic to a subgroup of the additive rationals $\mathbb{Q}$ or of the additive quotient group $\mathbb{Q}/\mathbb{Z}$.
Namely, the order of every locally cyclic group must be countable; and thus, it is worth mentioning that any subset of a locally cyclic Polish group $G$ is Borel.

Fix an index $i \ge 1$.
Since every finite subgroup of a locally cyclic group is cyclic, and since every cyclic group of order $k$ is isomorphic to $\mathbb{Z}/k\mathbb{Z}$ under addition, we observe that (i) every locally cyclic group $G$ has at most a countable number of finite normal subgroups $N \lhd G$; and (ii) $|Y_{i}| = k$ if and only if $Y_{i} \in G/N$ with $N \cong \mathbb{Z}/k\mathbb{Z}$.
Hence, it follows that $\mathbb{P}\{ |Y_{i}| = k \} = \mathbb{P}\{ Y_{i} \in G/N \ \mathrm{for} \ \mathrm{some} \ N \ \mathrm{isomorphic} \ \mathrm{to} \ \mathbb{Z}/k\mathbb{Z} \}$, and \lemref{lem:Q_Y} can be rewritten by
\begin{align}
H(X \mid Y)
=
\sum_{N \lhd G : \text{$N$ is finite}} (\log |N|) \, \mathbb{P}\{ Y_{i} \in G/N \} . 
\label{eq:formula_condH_LCG}
\end{align}
Namely, the probability $\mathbb{P}\{ Y_{i} \in G/N \}$ is important in the subsequent analysis.
Actually, if we define $\varepsilon_{n}^{(i)}(N) \coloneqq \mathbb{P}\{ Y_{n}^{(i)} \in G/N \}$ for each $i \ge 1$, $n \ge 0$, and finite $N \lhd G$, then it follows from \thref{th:erasure_recursive} that for every $i \ge 1$, $n \ge 1$ and finite $N \lhd G$, 
\begin{align}
\varepsilon_{n}^{(2i-1)}( N )
& =
\sum_{\substack{ H, K \lhd G : H \bullet K = N }} \varepsilon_{n-1}^{(i)}( H ) \, \varepsilon_{n-1}^{(i+2^{n-1})}( K ) ,
\label{eq:recursive_probab1} \\
\varepsilon_{n}^{(2i)}( N )
& =
\sum_{\substack{ H, K \lhd G : H \cap K = N }} \varepsilon_{n-1}^{(i)}( H ) \, \varepsilon_{n-1}^{(i+2^{n-1})}( K ) .
\label{eq:recursive_probab2} 
\end{align}
Based on the recursive formulas \eqref{eq:recursive_probab1}--\eqref{eq:recursive_probab2},
we observe from \corref{cor:erasure_recursive} and \eqref{eq:formula_condH_LCG} that for each $i \ge 1$ and $n \ge 0$,
\begin{align}
\mathfrak{H}_{n}^{(i)}
=
\sum_{N \lhd G : \text{$N$ is finite}} (\log |N|) \, \varepsilon_{n}^{(i)}( N ) .
\label{eq:erasure_recursive_condH}
\end{align}

\subsection{Asymptotic Distribution of Multilevel Polarization}
\label{sect:asymptotic_distribution}

In the previous subsection, the calculations of $\{ \mathfrak{H}_{n}^{(i)} \}_{i=1,}^{\infty} {}_{n=0}^{\infty}$ have been simplified to \eqref{eq:erasure_recursive_condH} via \eqref{eq:recursive_probab1}--\eqref{eq:recursive_probab2}.
Based on this, we now give a multilevel polarization theorem for erasure distributions with a locally cyclic group $G$.
To formalize it, we now introduce some notations and definitions as follows:

For each finite $N \lhd G$, denote by $\mathcal{S}( N )$ the collection of overgroups $H \rhd N$ satisfying the following two conditions: (i) there exists a $K \lhd G$ satisfying the proper subgroup chain $N \lhd K \lhd H$ and (ii) there are no distinct $K_{1}, K_{2} \lhd G$ satisfying the proper subgroup chain $N \lhd K_{1} \lhd K_{2} \lhd H$.
Moreover, for each finite $N \lhd G$ and $H \in \mathcal{S}(N)$, denote by $\mathcal{M}( N, H )$ the collection of finite subgroups $K \lhd G$ satisfying the proper subgroup chain $N \lhd K \lhd H$.
These notations will be used in Algorithm~\ref{alg:main} later.

For each finite $N \lhd G$, define
\begin{align}
Q( N )
\coloneqq
\lim_{m \to \infty} \frac{ 1 }{ m } \sum_{i = 1}^{m} \mathbb{P}\{ Y_{i} \in G/N \} ,
\label{def:Q}
\end{align}
provided that the limit exists.
Henceforth, assume that the limit $Q( N )$ always exists for each finite $N \lhd G$, and $Q( N ) \coloneqq 0$ if $N$ is infinite.
This existence of limits is a similar assumption to \cite[Remark~1]{alsan_telatar_2016} in the study of polar codes for non-stationary channels, which can be considered as the existence of entropy rate for a sequence of independent r.v.'s (see also \cite{mahdavifar_2016}).
The following example shows some simple cases of \eqref{def:Q}.

\begin{example}
If $\{ (X_{i}, Y_{i}) \}_{i = 1}^{\infty}$ is a stationary source with generic distribution $P_{X, Y}$, i.e., if $P_{X_{i}, Y_{i}} = P_{X, Y}$ for all $i \ge 1$, then it is clear that $Q( H ) = \mathbb{P}\{ Y \in G / H \}$.
Similarly, if $Y_{i}$ converges in distribution to $Y$ as $i \to \infty$, then it follows from the Ces\'{a}ro mean that $Q( H ) = \mathbb{P}\{ Y \in G / H \}$.
\end{example}

It is clear that $Q( \cdot )$ forms a discrete probability distribution on the set of normal subgroups of $G$.
As wil bel seen in \thref{th:asymptotic_distribution} and Algorithm~\ref{alg:main}, the distribution $Q( \cdot )$ plays a significant role to characterize the asymptotic distribution of multilevel source polarization.
For each $H, K, N \lhd G$, we define
\begin{align}
\chi(N, K, H)
& \coloneqq
\sum_{J \lhd G : J \bullet N = J \bullet K, J \cap H = J \cap K} Q( J ) ,
\label{def:chi_group} \\
\beta(N, H)
& \coloneqq
\sum_{J \lhd G : J \cap H = J \cap N} Q( J ) .
\label{def:beta_group}
\end{align}
The definitions \eqref{def:chi_group}--\eqref{def:beta_group} will be also used in Algorithm~\ref{alg:main}.

Finally, according to \eqref{def:mathfrak_H}, we define
\begin{align}
\mathfrak{H}_{n}^{(i)}[ N ]
\coloneqq
H( U_{n}^{(i)} \bullet N \mid \{ U_{n}^{(j)} \}_{j = 1}^{i-1}, \{ Y_{j} \}_{j = 1}^{\infty} )
\end{align}
for each $i \ge 1$, $n \ge 0$, and $N \lhd G$.
Since $U_{n}^{(i)} \bullet N$ is a function of $U_{n}^{(i)}$, it is clear that $\mathfrak{H}_{n}^{(i)}[ N ] \le \mathfrak{H}_{n}^{(i)}$ for each $N \lhd G$.
In addition, we readily see that $\mathfrak{H}_{n}^{(i)}[ G ] = 0$ and $\mathfrak{H}_{n}^{(i)}[ \{ e \} ] = \mathfrak{H}_{n}^{(i)}$, where $e \in G$ stands for the identity element of $G$.

The following theorem characterizes the asymptotic distribution of multilevel polarization for a non-stationary source following erasure distributions.

\begin{algorithm}[t]
\small
\DontPrintSemicolon
\KwData{A locally cyclic group $G$; a finite normal subgroup $N \lhd G$; and a distribution $Q( \cdot )$ defined in \eqref{def:Q}}
\KwResult{Probability masses $\{ \mu( H ) \}_{H \lhd N}$}
	$\alpha \longleftarrow 0$; $K \longleftarrow \{ e \}$; and $\mu( H ) \longleftarrow 0$ for all $H \lhd N$\;
	\While{$K \lhd N$}{
		\uIf{\emph{$\mathcal{S}( K )$ is nonempty}}{
			$(H_{1}, H_{2}) \longleftarrow \argmax\limits_{(K_{1}, K_{2}) : K_{1} \in \mathcal{S}( K ),  K_{2} \in \mathcal{M}(K, K_{1})} \chi(K, K_{2}, K_{1})$\hspace*{-2em}\;
			$\mu( K ) \longleftarrow \beta(K, H_{1}) - \alpha$\;
			\If{\emph{there exists $H_{3} \in \mathcal{M}(K, H_{1})$ s.t.\ $H_{3} \neq H_{2}$}}{
				$\mu( K ) \longleftarrow \mu( K ) + \chi(K, H_{3}, H_{1})$\;
			}
			$K \longleftarrow H_{2}$\;
		}
		\uElseIf{\emph{there exists an overgroup $H \rhd K$}}{
			$\mu( K ) \longleftarrow \beta(K, H) - \alpha$\;
			$K \longleftarrow H$\;
		}
		\Else{
			$\mu( K ) \longleftarrow 1 - \alpha$\;
		}
		$\alpha \longleftarrow \alpha + \mu( K )$\;
	}
\caption{Solving $\mu(N)$ used in \eqref{eq:asymptotic_distribution} of \thref{th:asymptotic_distribution}}
\label{alg:main}
\end{algorithm}

\begin{theorem}
\label{th:asymptotic_distribution}
For any $\delta > 0$ and finite $N \lhd G$, it holds that
\begin{align}
&
\lim_{n \to \infty} \lim_{m \to \infty} \frac{ 1 }{ m \, 2^{n} } \Big|\Big\{ 1 \le i \le m \, 2^{n} : \mathfrak{H}_{n}^{(i)}[N] < \delta ,
\notag \\
& \qquad \qquad \qquad \qquad \quad
\big| \mathfrak{H}_{n}^{(i)} - \log |N| \big| < \delta \Big\}\Big|
=
\mu(N) ,
\label{eq:asymptotic_distribution}
\end{align}
where $\mu( N )$ can be exactly calculated by Algorithm~\ref{alg:main}.
\end{theorem}

In \eqref{eq:asymptotic_distribution}, the limit with respect to $m$ is taken due to the non-stationarity of the source and the dependence of r.v.'s induced by the polar transforms \eqref{eq:polar_transform_minus}--\eqref{eq:polar_transform_plus}, see \sectref{sect:recursive}.
Here, note that the number $m$ plays a similar role to that of \sectref{sect:recursive}; however, it is not given as $m \coloneqq \lfloor i / 2^{n} \rfloor$ but given as independent of $n$ in \thref{th:asymptotic_distribution}.

We shall prove \thref{th:asymptotic_distribution} by employing elementary techniques in \emph{lattice theory} \cite{birkhoff_1967}.
Basic notions and definitions in lattice theory can be found in \appref{app:lattice}.

\begin{IEEEproof}[Proof of \thref{th:asymptotic_distribution}]
Since the lattice of normal subgroups of a group is distributive if and only if the group is locally cyclic (cf.\ \cite[Theorem~4 of Chapter~3]{ore_1938}), it suffices to consider a distributive lattice $(L, \vee, \wedge, \le)$.
Define the double sequence $\{ \bvec{\varepsilon}_{n}^{(i)} \}_{i = 1,}^{\infty} {}_{n = 0}^{\infty}$ of probability vectors $\bvec{\varepsilon}_{n}^{(i)} \coloneqq \{ \varepsilon_{n}^{(i)}( j ) \}_{j \in L}$ by
\begin{align}
\varepsilon_{n}^{(2 i - 1)}( j )
& \coloneqq
\sum_{k, l \in L : k \vee l = j} \varepsilon_{n-1}^{(i)}( k ) \, \varepsilon_{n-1}^{(i+2^{n-1})}( l ) ,
\label{def:recursive_eps_lattice1} \\
\varepsilon_{n}^{(2 i)}( j )
& \coloneqq
\sum_{k, l \in L : k \wedge l = j} \varepsilon_{n-1}^{(i)}( k ) \, \varepsilon_{n-1}^{(i+2^{n-1})}( l )
\label{def:recursive_eps_lattice2}
\end{align}
for each $i \ge 1$ and $n \ge 1$, where $\{ \bvec{\varepsilon}_{0}^{(i)} \}_{i = 1}^{\infty} \coloneqq \{ \bvec{\varepsilon}_{i} \}_{i = 1}^{\infty}$ is a given initial probability vector.
Note that \eqref{def:recursive_eps_lattice1} and \eqref{def:recursive_eps_lattice2} correspond to \eqref{eq:recursive_probab1} and \eqref{eq:recursive_probab2}, respectively.
To analyze the probability vectors $\bvec{\varepsilon}_{n}^{(i)}$, we further define partial sums of elements in $\bvec{\varepsilon}_{n}^{(i)}$ by
\begin{align}
\theta_{n}^{(i)}(a, b)
& \coloneqq
\sum_{ j \in L : j \vee a = j \vee b} \varepsilon_{n}^{(i)}( j ) ,
\label{def:theta} \\
\chi_{n}^{(i)}(a, c, b)
& \coloneqq
\sum_{ j \in L : j \vee a = j \vee c , j \wedge b = j \wedge c } \varepsilon_{n}^{(i)}( j ) ,
\label{def:chi} \\
\beta_{n}^{(i)}(a, b)
& \coloneqq
\sum_{ j \in L : j \wedge a = j \wedge b } \varepsilon_{n}^{(i)}( j ) 
\label{def:beta}
\end{align}
for each $a, b, c \in L$.
When elements $a, b \in L$ are clear from the context, we simply write \eqref{def:theta}--\eqref{def:beta} as $\theta_{n}^{(i)}$, $\chi_{n}^{(i)}( c )$, and $\beta_{n}^{(i)}$.
By defining 
\begin{align}
\mathcal{M}(a, b)
\coloneqq
\{ c \in L \mid a < c < b \}
\label{def:M}
\end{align}
for each $a, b \in L$, we can show the following lemma.

\begin{lemma}
\label{lem:sum_is_unity}
Let $a, b \in L$ be chosen so that $a < b$ and there is no pair $x, y \in L$ satisfying $a < x < y < b$.
If $(L, \le)$ is modular, then it holds that
\begin{align}
\theta_{n}^{(i)} + \beta_{n}^{(i)} + \sum_{c \in \mathcal{M}(a, b)} \chi_{n}^{(i)}(c)
=
1
\end{align}
for every $a, b \in L$, $i \ge 1$, and $n \ge 0$.
\end{lemma}

\begin{IEEEproof}[Proof of \lemref{lem:sum_is_unity}]
See \appref{app:sum_is_unity}.
\end{IEEEproof}

Note that every distributive lattice is modular.
\lemref{lem:sum_is_unity} means that the probability masses of $\bvec{\varepsilon}_{n}^{(i)}$ are well-partitioned by \eqref{def:theta}--\eqref{def:beta}.
Let $a$ and $b$ be chosen so that $a < b$ and there is no pair $x, y \in L$ satisfying $a < x < y < b$.
If $(L, \le)$ is modular, then $|\mathcal{M}(a, b)|$ can be an arbitrary nonnegative integer.
Particularly, if $(L, \le)$ is distributive, then it can be verified that $0 \le |M(a, b)| \le 2$.
Noting this fact, we can observe:

\begin{lemma}
\label{lem:formulae}
Let $(L, \le)$ be a distributive lattice, and let $a, b \in L$ be chosen so that $a < b$.
Suppose that there is no pair $x, y \in L$ satisfying $a < x < y < b$.
Then, it holds that
\begin{align}
\theta_{n}^{(2i-1)}
& =
\theta_{n-1}^{(i)} + \theta_{n-1}^{(i+2^{n-1})} - \theta_{n-1}^{(i)} \, \theta_{n-1}^{(i+2^{n-1})} + C_{n}^{(i)} ,
\\
\beta_{n}^{(2i-1)}
& =
\beta_{n-1}^{(i)} \, \beta_{n-1}^{(i+2^{n-1})} ,
\\
\theta_{n}^{(2i)}
& =
\theta_{n-1}^{(i)} \, \theta_{n-1}^{(i+2^{n-1})} ,
\\
\beta_{n}^{(2i)}
& =
\beta_{n-1}^{(i)} + \beta_{n-1}^{(i+2^{n-1})} - \beta_{n-1}^{(i)} \, \beta_{n-1}^{(i+2^{n-1})} + C_{n}^{(i)}
\end{align}
for every $n \ge 1$ and $i \ge 1$, where
\begin{align}
C_{n}^{(i)}
& \coloneqq
\sum_{c_{1}, c_{2} \in \mathcal{M}(a, b) : c_{1} \neq c_{2}} \chi_{n-1}^{(i)}( c_{1} ) \, \chi_{n-1}^{(i+2^{n-1})}( c_{2} ) .
\label{def:Cni}
\end{align}
Moreover, if there exists an $x \in L$ satisfying $a \prec x \prec b$, then it holds that
\begin{align}
\chi_{n}^{(2 i - 1)}(c)
& =
\chi_{n-1}^{(i)}(c) \, \chi_{n-1}^{(i+2^{n-1})}(c)
\notag \\
& \quad
{} + \chi_{n-1}^{(i)}(c) \, \beta_{n-1}^{(i+2^{n-1})} + \beta_{n-1}^{(i)} \, \chi_{n-1}^{(i+2^{n-1})}(c) ,
\\
\chi_{n}^{(2 i)}(c)
& =
\chi_{n-1}^{(i)}(c) \, \chi_{n-1}^{(i+2^{n-1})}(c)
\notag \\
& \quad
{} + \chi_{n-1}^{(i)}(c) \, \theta_{n-1}^{(i+2^{n-1})} + \theta_{n-1}^{(i)} \, \chi_{n-1}^{(i+2^{n-1})}(c)
\end{align}
for every $n \ge 1$, every $i \ge 1$ and every $c \in \mathcal{M}(a, b)$.
\end{lemma}

\begin{IEEEproof}[Proof of \lemref{lem:formulae}]
See \appref{app:formulae}.
\end{IEEEproof}

Since the positive divisors of a positive integer form a distributive lattice, \lemref{lem:formulae} is a generalization of \cite[Lemma~6]{sakai_iwata_fujisaki_2018} from a lattice of positive divisors with the stationary source setting to general distributive lattices with the non-stationary source setting.
Therefore, as in \cite[Lemma~7]{sakai_iwata_fujisaki_2018}, we can obtain the following lemma.

\begin{lemma}
\label{eq:sub_super_martingale}
Let $(L, \le)$ be a distributive lattice, and let $a, b \in L$ be chosen so that $a < b$.
Suppose that there is no pair $x, y \in L$ satisfying $a < x < y < b$.
Then, it holds that
\begin{align}
\theta_{n}^{(2i-1)} + \theta_{n}^{(2i)}
& =
\theta_{n-1}^{(i)} + \theta_{n-1}^{(i+2^{n-1})} + C_{n}^{(i)} ,
\\
\beta_{n}^{(2i-1)} + \beta_{n}^{(2i)}
& =
\beta_{n-1}^{(i)} + \beta_{n-1}^{(i+2^{n-1})} + C_{n}^{(i)}
\end{align}
for every $n \ge 1$ and $i \ge 1$.
Moreover, if there exists an $x \in L$ satisfying $a \prec x \prec b$, then it holds that
\begin{align}
\chi_{n}^{(2i-1)}( c ) + \chi_{n}^{(2i)}( c )
& =
\chi_{n-1}^{(i)} \, \Big( 1 - \chi_{n-1}^{(i+2^{n-1})} \Big)
\notag \\
& \qquad
{} + \chi_{n-1}^{(i+2^{n-1})} \, \Big( 1 - \chi_{n-1}^{(i)} \Big)
\end{align}
\end{lemma}

\begin{IEEEproof}[Proof of \lemref{eq:sub_super_martingale}]
\lemref{eq:sub_super_martingale} is a direct consequence from Lemmas~\ref{lem:sum_is_unity} and~\ref{lem:formulae}.
\end{IEEEproof}

Based on this observation, we can prove \thref{th:asymptotic_distribution} in a similar fashion to \cite[Section~IV]{sakai_iwata_fujisaki_2018}.
\end{IEEEproof}

\section{Concluding Remarks}

We have explored the asymptotic distribution of multilevel source polarization over possibly infinite source alphabets by defining a convenient probabilistic model called an erasure distribution, which is defined in \defref{def:group_erasure}.
The analysis of Ar{\i}kan-style two-by-two polar transforms \eqref{eq:polar_transform_minus}--\eqref{eq:polar_transform_plus} based on a Polish group was simplified by \thref{th:erasure_recursive} establishing recursive formulas of the polar transforms for erasure distributions.
When the group is locally cyclic, \thref{th:asymptotic_distribution} and Algorithm~\ref{alg:main} give a method for calculating the exact asymptotic distribution of multilevel source polarization for erasure distributions.
This is the first instance of multilevel source polarization with countably infinite levels, which is characterized by the structure of distributive lattices.

\subsection{Simple Instances of \thref{th:asymptotic_distribution}}

In the following, we mention two examples of \thref{th:asymptotic_distribution}.

\subsubsection{Modular Arithmetic Erasure Channels}
\label{sect:MAEC}

As explained in Remarks~\ref{rem:reduction} and~\ref{rem:reduction2}, the erasure distribution defined in \defref{def:group_erasure} can be reduced to the \emph{modular arithmetic erasure channel} \cite[Definition~2]{sakai_iwata_fujisaki_2018}, provided that $G$ is a finite cyclic group.
Since every cyclic group is locally cyclic, \thref{th:asymptotic_distribution} can also be reduced to the authors' previous results \cite{sakai_iwata_fujisaki_isit2018, sakai_iwata_fujisaki_2018}, which is described in a stationary setting.

\subsubsection{Pr\"{u}fer $p$-group}
\label{sect:prufer}

Let $p \ge 2$ be a prime number.
The \emph{Pr\"{u}fer $p$-group} $G$ can be defined by the Sylow $p$-subgroup of $\mathbb{Q}/\mathbb{Z}$ up to isomorphism, i.e., $G \simeq \{ m/p^{n} + \mathbb{Z}\mid m \in \mathbb{Z} \ \mathrm{and} \ n \in \mathbb{Z}_{\ge 0} \}$.
It is known that $H \lhd K$ or $K \lhd H$ for any $H, K \lhd G$, provided that $G$ is the Pr\"{u}fer $p$-group.
Thus, \corref{cor:asymptotic_distribution_Prufer} can simplify \thref{th:asymptotic_distribution} without Algorithm~\ref{alg:main}.

\begin{corollary}
\label{cor:asymptotic_distribution_Prufer}
Suppose that $G$ is the Pr\"{u}fer $p$-group.
For any $\delta > 0$ and finite $N \lhd G$, and $Q( \cdot )$ as in \eqref{def:Q}, it holds that
\begin{align}
&
\lim_{n \to \infty} \lim_{m \to \infty} \frac{ 1 }{ m \, 2^{n} } \Big|\Big\{ 1 \le i \le m \, 2^{n} : \mathfrak{H}_{n}^{(i)}[N] < \delta ,
\notag \\
& \qquad \qquad \qquad \qquad \quad
\big| \mathfrak{H}_{n}^{(i)} - \log |N| \big| < \delta \Big\}\Big|
=
Q(N) .
\end{align}
\end{corollary}

Therefore, the source polarization for erasure distributions can be simply characterized by the initial condition \eqref{def:Q} without any other compulation method like Algorithm~\ref{alg:main}, provided that $G$ is the Pr\"{u}fer $p$-group.
This gives a simple and concrete instance of countably infinite polarization levels.

\subsection{Future Works}

We have shown a possibility of multilevel polarization phenomena over infinite alphabets.
Multilevel polarization analysis for more general source distributions than erasure distributions is an open problem;
and inventing practical encoding/decoding schemes with the infinite polarization levels is also of interest in the study of source coding over an infinite source alphabet.

While \thref{th:asymptotic_distribution} can be reduced to the authors' previous results \cite{sakai_iwata_fujisaki_2018, sakai_iwata_fujisaki_isit2018} as discussed in \sectref{sect:MAEC}, it cannot be reduced to Nasser--Telatar's case study \cite[Section~VIII]{quasigroup}, because an elementary abelian group is not locally cyclic in general (cf.\ Remarks~\ref{rem:reduction} and~\ref{rem:reduction2}).
Generalizing \thref{th:asymptotic_distribution} from locally cyclic to abelian, or not necessarily abelian, groups is highly of interest in terms of the MAC polarization \cite{quasigroup, fourier_analysis}.

\section*{Acknowledgements}

The authors would like to thank to Dr.~Jun Muramatsu for his helpful discussions; Dr.~Mine Alsan for her valuable comments greatly improving this paper; and the anonymous reviewers in ISIT'19 for carefully reading this paper and for giving many valuable advices.

\appendices

\section{Brief Introduction to Lattice Theory}
\label{app:lattice}

\begin{definition}[partially ordered sets; posets]
For a binary relation $\le$ on a nonempty set $L$, the system $(L, \le)$ is called a \emph{poset} if it satisfies the following three properties: (i) $a \le a$; (ii) $a \le b$ and $b \le a$ imply that $a = b$;
and (iii) $a \le b$ and $b \le c$ imply that $a \le c$, for all $a, b, c \in L$.
\end{definition}

Let $(L, \le)$ be a poset.
As a strict relation, the binary relation $a < b$ is a shorthand for $a \le b$ and $a \neq b$.

\begin{definition}[predecessors and followers]
For two elements $a, b \in L$ of a poset $(L, \le)$, we say that \emph{$b$ covers $a$,} or \emph{$a$ is covered by $b$,} if $a < b$ and there is no $x \in L$ such that $a < x < b$.
This relation is denoted by $a \prec b$.
\end{definition}

For each $a, b \in L$, an \emph{upper bound} of $a$ and $b$ is an element $u \in L$ satisfying $a \le u$ and $b \le u$; and a \emph{least upper bound} $s$ of $a$ and $b$ is an upper bound of $a$ and $b$ satisfying $s \le u$ for every upper bound $u$ of $a$ and $b$.
If a least upper bound $s$ of $a$ and $b$ exists, then it is unique; and thus, it can be denoted by $a \vee b \coloneqq s$, provided that it exists.
Analogously, for each $a, b \in L$, a \emph{lower bound} of $a$ and $b$ is an element $l \in L$ satisfying $l \le a$ and $l \le b$; and a \emph{greatest lower bound} $i$ of $a$ and $b$ is an upper bound of $a$ and $b$ satisfying $l \le i$ for every lower bound $i$ of $a$ and $b$.
If a greatest lower bound $i$ of $a$ and $b$ exists, then it is unique; and thus, it can be denoted by $a \wedge b \coloneqq i$, provided that it exists.

\begin{definition}[lattices]
A poset $(L, \le)$ is called a \emph{lattice} if every two elements $a, b \in L$ have the least upper bound $a \vee b$ and the greatest lower bound $a \wedge b$.
\end{definition}

Given a lattice $(L, \le)$, the binary operations $\vee$ and $\wedge$ are called a \emph{join} and a \emph{meet}, respectively, and the lattice is sometimes denoted by $(L, \vee, \wedge, \le)$.
These binary operations $\vee$ and $\wedge$ satisfy the following identities:

\begin{lemma}[{\cite[Lemma~1 in page~8]{birkhoff_1967}}]
Let $(L, \le)$ be a lattice.
For every $a, b, c \in L$, it holds that
\begin{itemize}
\item[(i)]
$a \vee b = b \vee a$ and $a \wedge b = b \wedge a$;
\item[(ii)]
$a \vee (b \vee c) = (a \vee b) \vee c$ and $a \wedge (b \wedge c) = (a \wedge b) \wedge c$; and
\item[(iii)]
$a \vee (a \wedge b) = a \wedge (a \vee b) = a$.
\end{itemize}
\end{lemma}

We now give the notion of modularity as follows:

\begin{definition}[modular lattices; {\cite[Section~7 of Chapter~I]{birkhoff_1967}}]
\label{def:modular}
A lattice $(L, \le)$ is said to be \emph{modular} if
$a \le c$ implies that $a \vee (b \wedge c) = (a \vee b) \wedge c$ for every $a, b, c \in L$.
\end{definition}

We readily see that \defref{def:modular} is equivalent to the following two identities:
\begin{align}
[(x \wedge y) \vee z] \wedge y
& =
(x \wedge y) \vee (z \wedge y) ,
\label{id:modular1} \\
[(x \vee y) \wedge z] \vee y
& =
(x \vee y) \wedge (z \vee y) .
\label{id:modular2} 
\end{align}

We next give the notion of distributivity as follows:

\begin{definition}[distributive lattices; {\cite[Section~6 of Chapter~I]{birkhoff_1967}}]
A lattice $(L, \le)$ is said to be \emph{distributive} if
\begin{align}
a \vee (b \wedge c)
& =
(a \vee b) \wedge (a \vee c) ,
\label{def:distributive1} \\
a \wedge (b \vee c)
& =
(a \wedge b) \vee (a \wedge c)
\label{def:distributive2}
\end{align}
for every $a, b, c \in L$.
\end{definition}

Note that every distributive lattice is modular, but there is a modular lattice which is not distributive.
Finally, we give the following lemma under the distributivity.

\begin{lemma}
\label{lem:distributive}
Let $(L, \le)$ be a distributive lattice, and let $a, b \in L$ be chosen so that $a \le b$.
Then, it holds that
\begin{align}
j \wedge a
=
j \wedge b
\ \mathrm{and} \
k \wedge a
=
k \wedge b
\end{align}
if and only if
\begin{align}
(j \vee k) \wedge a
=
(j \vee k) \wedge b ,
\end{align}
for every $j, k \in L$.
\end{lemma}

\begin{IEEEproof}[Proof of \lemref{lem:distributive}]
\subsubsection{If part $\Leftarrow$}
We readily see that
\begin{align}
j \wedge a
& =
[(j \vee k) \wedge j] \wedge a
\notag \\
& =
[(j \vee k) \wedge a] \wedge j
\notag \\
& =
[(j \vee k) \wedge b] \wedge j
\notag \\
& =
[(j \vee k) \wedge j] \wedge b
\notag \\
& =
j \wedge b .
\end{align}
The identity $k \wedge a = k \wedge b$ can be shown similarly; and these imply the sufficiency, as desired.

\subsubsection{Only if part $\Rightarrow$}
We readily see that
\begin{align}
(j \vee k) \wedge a
& =
(j \wedge a) \vee (k \wedge a)
\notag \\
& =
(j \wedge b) \vee (k \wedge b)
\notag \\
& =
(j \vee k) \wedge b ,
\end{align}
which implies the necessity, as desired.
\end{IEEEproof}

While \lemref{lem:distributive} is elementary, this lemma is important in proving \thref{th:asymptotic_distribution}.
Note that \lemref{lem:distributive} does not hold, provided that a lattice $(L, \le)$ is modular but not distributive.

\section{Proof of \lemref{lem:sum_is_unity}}
\label{app:sum_is_unity}

Let $a, b \in L$ be chosen so that $a < b$, and let $i \in L$ be an arbitrary element.
Since $a \wedge b = a < b = a \vee b$, note that \textcolor{black}{$i \vee a \le i \vee b$ and $i \wedge a \le i \wedge b$}.
If $i \vee a = i \vee b$, then
\begin{align}
a
& =
(i \wedge a) \vee a
\notag \\
& \le
(i \wedge b) \vee a
\notag \\
& \overset{\mathclap{\text{\textcolor{black}{(a)}}}}{=}
(i \vee a) \wedge b
\notag \\
& =
(i \vee b) \wedge b
\notag \\
& =
b ,
\end{align}
\textcolor{black}{where (a) follows by the modularity (see \defref{def:modular}).
As $a < b$,} this implies that $i \wedge a \neq i \wedge b$;
and therefore, since $i \wedge a \le i \wedge b$, it follows that $i \wedge a < i \wedge b$ if $i \vee a = i \vee b$.
Similarly, it can be dually verified that $i \vee a < i \vee b$ if $i \wedge a = i \wedge b$.
Therefore, we have
\begin{align}
\{ j \in L \mid j \vee a = j \vee b \} \cap \{ k \in L \mid k \wedge a = k \wedge b \}
=
\emptyset .
\end{align}
On the other hand, if $i \vee a < i \vee b$ and $i \wedge a < i \wedge b$, then it holds that \textcolor{black}{$a < (i \wedge b) \vee a < b$}, which implies the existence of $x \in L$ satisfying $a < x < b$.

We first consider the case where there is no $x \in L$ satisfying $a < x < b$, i.e., suppose that $a \prec b$.
Then, the set $\mathcal{M}(a, b)$ defined in \eqref{def:M} is empty, and we observe that
\begin{align}
\{ \{ j \in L \mid j \vee a = j \vee b \}, \{ k \in L \mid k \wedge a = k \wedge b \} \}
\end{align}
forms a partition of $L$.
Therefore, since $\bvec{\varepsilon}_{n}^{(i)}$ is a probability vector for each $n \ge 0$ and $i \ge 1$, it follows from \eqref{def:theta}--\eqref{def:beta} that \lemref{lem:sum_is_unity} holds, provided that $a \prec b$.

We next consider the case where there is at least one $x \in L$ such that \textcolor{black}{$a \prec x \prec b$}.
Then, the set $\mathcal{M}(a, b)$ defined in \eqref{def:M} is nonempty.
Moreover, it follows by the modularity of \defref{def:modular} that $a \prec y \prec b$ for every $y \in \mathcal{M}(a, b)$.
Hence, we observe that neither $c_{1} \le c_{2}$ nor $c_{2} \le c_{1}$ for every distinct $c_{1}, c_{2} \in \mathcal{M}(a, b)$.
If $i \vee a = i \vee b$, then it is clear that $i \vee c = i \vee b$ for every $c \in \mathcal{M}(a, b)$; and it follows that
\begin{align}
a
& =
(i \wedge a) \vee a
\notag \\
& \le
(i \wedge c) \vee a
\notag \\
& =
(i \vee a) \wedge c
\notag \\
& =
(i \vee c) \wedge c
\notag \\
& =
c
\end{align}
for every $c \in \mathcal{M}(a, b)$, which implies that $i \wedge a < i \wedge c$ for every $c \in \mathcal{M}(a, b)$.
Similarly, if $i \wedge a = i \wedge b$, then it can be dually verified that $i \vee c < i \vee b$ and $i \wedge c = i \wedge a$ for every $c \in \mathcal{M}(a, b)$.
Now, suppose that $i \vee a < i \vee b$ and $i \wedge a < i \wedge b$.
As shown in the two paragraphs back, there exists $c \in \mathcal{M}(a, b)$ such that
\begin{align}
a < c = (i \wedge b) \vee a < b .
\end{align}
For such an element $c = (i \wedge b) \vee a = (i \vee a) \wedge b$, it holds that
\begin{align}
i \wedge c
& =
i \wedge [(i \wedge b) \vee a]
\notag \\
& \overset{\mathclap{\text{(a)}}}{=}
(i \wedge b) \vee (i \wedge a)
\notag \\
& = i \wedge b
\end{align}
and
\begin{align}
i \vee c
& =
i \vee [(i \vee a) \wedge b]
\notag \\
& \overset{\mathclap{\text{(b)}}}{=}
(i \vee a) \wedge (i \vee b)
\notag \\
& =
i \vee a ,
\end{align}
where (a) and (b) follow from \eqref{id:modular1} and \eqref{id:modular2}, respectively.
For each $c \in \mathcal{M}(a, b)$, define
\begin{align}
S( c )
\coloneqq
\{ x \in L \mid x \vee a = x \vee c \ \mathrm{and} \ x \wedge b = x \wedge c \} .
\end{align}
\textcolor{black}{We now prove by contradiction that $S( c_{1} ) \cap S( c_{2} ) = \emptyset$ for every distinct $c_{1}, c_{2} \in \mathcal{M}(a, b)$.
That is, suppose that there exists an element $d \in L$ satisfying $d \in S( c_{1} ) \cap S( c_{2} )$.
Then, we observe that}
\begin{align}
c_{1}
& \overset{\mathclap{\text{(a)}}}{=}
c_{1} \wedge b
\notag \\
& \overset{\mathclap{\text{(b)}}}{\le}
(d \vee c_{1}) \wedge b
\notag \\
& \overset{\mathclap{\text{(c)}}}{=}
(d \vee c_{2}) \wedge b
\notag \\
& \overset{\mathclap{\text{(d)}}}{=}
(d \wedge b) \vee c_{2}
\notag \\
& \overset{\mathclap{\text{(e)}}}{=}
(d \wedge c_{2}) \vee c_{2}
\notag \\
& \overset{\mathclap{\text{(f)}}}{=}
c_{2} ,
\end{align}
where
(a) follows from $c_{1} \prec b$;
(b) follows from $c_{1} \le d \vee c_{1}$;
(c) follows from $d \in S( c_{1} ) \cap S( c_{2} )$, i.e., $d \vee c_{1} = d \vee a = d \vee c_{2}$;
(d) follows from $c_{2} \le b$ and the modular equality \eqref{def:modular};
(e) follows from $d \in S( c_{1} ) \cap S( c_{2} )$, i.e., $d \wedge c_{1} = d \wedge b = d \wedge c_{2}$; and
(f) follows from the absorption law: $(x \wedge y) \vee y = y$.
This, however, contradicts to $c_{1} \not\le c_{2}$.
Therefore, we have $S( c_{1} ) \cap S( c_{2} ) = \emptyset$.
Concluding discussions of this paragraph, we observe that
\begin{align}
\emptyset
& =
\{ j \in L \mid j \vee a = j \vee b \} \cap \{ k \in L \mid k \wedge a = k \wedge b \}
\\
& =
\{ j \in L \mid j \vee a = j \vee b \} \cap S( c )
\\
& =
\{ k \in L \mid k \wedge a = k \wedge b \} \cap S( c )
\\
& =
S( c_{1} ) \cap S( c_{2} )
\end{align}
for every $c \in \mathcal{M}(a, b)$ and every distinct $c_{1}, c_{2} \in \mathcal{M}(a, b)$; and
\begin{align}
&
\bigcup_{c \in \mathcal{M}(a, b)} S( c )
\notag \\
& \quad =
\Big( \{ j \in L \mid j \vee a = j \vee b \} \cup \{ k \in L \mid k \wedge a = k \wedge b \} \Big)^{\complement} ,
\end{align}
where $\mathcal{A}^{\complement}$ stands for the complement of a set $\mathcal{A}$.
Thus, we have that
\begin{align}
&
\{ \{ j \in L \mid j \vee a = j \vee b \}, \{ k \in L \mid k \wedge a = k \wedge b \} \}
\notag \\
& \qquad \qquad \qquad \qquad \qquad \qquad \cup
\{ S( c ) \mid c \in \mathcal{M}(a, b) \}
\end{align}
forms a partition of $L$.
Therefore, since $\bvec{\varepsilon}_{n}^{(i)}$ is a probability vector for each $n \ge 0$ and $i \ge 1$, it follows from \eqref{def:theta}--\eqref{def:beta} that \lemref{lem:sum_is_unity} holds, provided that there exists an $x \in L$ satisfying $a \prec x \prec b$.
This completes the proof of \lemref{lem:sum_is_unity}.
\hfill\IEEEQEDhere

\section{Proof of \lemref{lem:formulae}}
\label{app:formulae}

By the duality between the join $\vee$ and the meet $\wedge$, it suffices to prove the identities of \lemref{lem:formulae} only for the minus transforms \eqref{def:recursive_eps_lattice1}.
Let $a, b \in L$ be chosen so that $a < b$.
A direct calculation shows
\begin{align}
\beta_{n}^{(2i-1)} \,
& \overset{\mathclap{\text{(a)}}}{=}
\sum_{\substack{ j \in L : \\ j \wedge a = j \wedge b }} \varepsilon_{n}^{(2i-1)}( j ) 
\notag \\
& \overset{\mathclap{\text{(b)}}}{=}
\sum_{\substack{ j \in L : \\ j \wedge a = j \wedge b }} \sum_{\substack{ k, l \in L : \\ k \vee l = j }} \varepsilon_{n-1}^{(i)}( k ) \, \varepsilon_{n-1}^{(i+2^{n-1})}( l )
\notag \\
& =
\sum_{\substack{ k, l \in L : \\ (k \vee l) \wedge a = (k \vee l) \wedge b }} \varepsilon_{n-1}^{(i)}( k ) \, \varepsilon_{n-1}^{(i+2^{n-1})}( l )
\notag \\
& \overset{\mathclap{\text{(c)}}}{=}
\sum_{\substack{ k \in L : \\ k \wedge a = k \wedge b }} \varepsilon_{n-1}^{(i)}( k ) \sum_{\substack{ l \in L : \\ l \wedge a = l \wedge b }} \varepsilon_{n-1}^{(i+2^{n-1})}( l )
\notag \\
& =
\beta_{n-1}^{(i)} \, \beta_{n-1}^{(i+2^{n-1})}
\label{eq:beta_minus}
\end{align}
for every $n \ge 1$ and $i \ge 1$, where
(a) follows from \eqref{def:beta};
(b) follows from \eqref{def:recursive_eps_lattice1}; and
(c) follows from \lemref{lem:distributive}.

Suppose that there is at least one $x \in L$ such that $a \prec x \prec b$, i.e., $b$ covers $x$ and $x$ covers $a$.
For each $c \in \mathcal{M}(a, b)$, we have
\begin{align}
\chi_{n}^{(2 i - 1)}(c) \,
& \overset{\mathclap{\text{(a)}}}{=}
\sum_{\substack{ j \in L : \\ j \vee a = j \vee c , \\ j \wedge b = j \wedge c }} \varepsilon_{n}^{(2i - 1)}( j ) 
\notag \\
& \overset{\mathclap{\text{(b)}}}{=}
\sum_{\substack{ i \in L : \\ i \vee c = i \vee a , \\ i \wedge c = i \wedge b \; }} \sum_{\substack{ k, l \in L : \\ k \vee l = j }} \varepsilon_{n-1}^{(i)}( k ) \, \varepsilon_{n-1}^{(i+2^{n-1})}( l )
\notag \\
& =
\sum_{\substack{ k, l \in L : \\ (k \vee l) \vee c = (k \vee l) \vee a , \\ (k \vee l) \wedge c = (k \vee l) \wedge b \; }} \varepsilon_{n-1}^{(i)}( k ) \, \varepsilon_{n-1}^{(i+2^{n-1})}( l )
\notag \\
& =
\sum_{\substack{ k, l \in L : \\ (k \vee l) \vee c = (k \vee l) \vee a , \\ (k \vee l) \wedge c = (k \vee l) \wedge b , \\ k \vee a = k \vee b , \\ l \vee a < l \vee b \; }} \varepsilon_{n-1}^{(i)}( k ) \, \varepsilon_{n-1}^{(i+2^{n-1})}( l )
\notag \\
& \qquad
{} + \sum_{\substack{ k, l \in L : \\ (k \vee l) \vee c = (k \vee l) \vee a , \\ (k \vee l) \wedge c = (k \vee l) \wedge b , \\ k \vee a < k \vee b , \\ l \vee a = l \vee b \; }} \varepsilon_{n-1}^{(i)}( k ) \, \varepsilon_{n-1}^{(i+2^{n-1})}( l )
\notag \\
& \qquad \qquad
{} + \sum_{\substack{ k, l \in L : \\ (k \vee l) \vee c = (k \vee l) \vee a , \\ (k \vee l) \wedge c = (k \vee l) \wedge b , \\ k \vee a < k \vee b , \\ l \vee a < l \vee b \; }} \varepsilon_{n-1}^{(i)}( k ) \, \varepsilon_{n-1}^{(i+2^{n-1})}( l )
\notag \\
& \overset{\mathclap{\text{(c)}}}{=}
\sum_{\substack{ k, l \in L : \\ (k \vee l) \wedge c = (k \vee l) \wedge b , \\ k \vee a = k \vee b , \\ l \vee a < l \vee b \; }} \varepsilon_{n-1}^{(i)}( k ) \, \varepsilon_{n-1}^{(i+2^{n-1})}( l )
\notag \\
& \qquad
{} + \sum_{\substack{ k, l \in L : \\ (k \vee l) \wedge c = (k \vee l) \wedge b , \\ k \vee a < k \vee b , \\ l \vee a = l \vee b \; }} \varepsilon_{n-1}^{(i)}( k ) \, \varepsilon_{n-1}^{(i+2^{n-1})}( l )
\notag \\
& \qquad \qquad
{} + \sum_{\substack{ j, k \in L : \\ (j \vee k) \vee c = (j \vee k) \vee a , \\ (j \vee k) \wedge c = (j \vee k) \wedge b , \\ j \vee a < j \vee b , \\ k \vee a < k \vee b \; }} \varepsilon_{n-1}^{(i)}( k ) \, \varepsilon_{n-1}^{(i+2^{n-1})}( l )
\notag \\
& \overset{\mathclap{\text{(d)}}}{=}
\sum_{\substack{ k, l \in L : \\ k \wedge c = k \wedge b , \\ l \wedge c = l \wedge b , \\ k \vee a = k \vee b , \\ l \vee a < l \vee b \; }} \varepsilon_{n-1}^{(i)}( k ) \, \varepsilon_{n-1}^{(i+2^{n-1})}( l )
\notag \\
& \qquad
{} + \sum_{\substack{ k, l \in L : \\ k \wedge c = k \wedge b , \\ l \wedge c = l \wedge b , \\ k \vee a < k \vee b , \\ l \vee a = l \vee b \; }} \varepsilon_{n-1}^{(i)}( k ) \, \varepsilon_{n-1}^{(i+2^{n-1})}( l )
\notag \\
& \qquad \qquad
{} + \sum_{\substack{ k, l \in L : \\ (k \vee l) \vee c = (k \vee l) \vee a , \\ k \wedge c = k \wedge b , \\ l \wedge c = l \wedge b , \\ k \vee a < k \vee b , \\ l \vee a < l \vee b \; }} \varepsilon_{n-1}^{(i)}( k ) \, \varepsilon_{n-1}^{(i+2^{n-1})}( l )
\notag \\
& \overset{\mathclap{\text{(e)}}}{=}
\sum_{\substack{ k, l \in L : \\ l \wedge a = l \wedge b , \\ k \vee c = k \vee b , \\ k \vee c = k \vee a \; }} \varepsilon_{n-1}^{(i)}( k ) \, \varepsilon_{n-1}^{(i+2^{n-1})}( l )
\notag \\
& \qquad
{} + \sum_{\substack{ k, l \in L : \\ k \wedge a = k \wedge b , \\ l \wedge c = l \wedge a , \\ l \vee c = l \vee b \; }} \varepsilon_{n-1}^{(i)}( k ) \, \varepsilon_{n-1}^{(i+2^{n-1})}( l )
\notag \\
& \qquad \qquad
{} + \sum_{\substack{ k, l \in L : \\ k \wedge c = k \wedge b , \\ l \wedge c = l \wedge b , \\ k \vee c = k \vee a , \\ l \vee c = l \vee a \; }} \varepsilon_{n-1}^{(i)}( k ) \, \varepsilon_{n-1}^{(i+2^{n-1})}( l )
\notag \\
& =
\beta_{n-1}^{(i)} \, \chi_{n-1}^{(i+2^{n-1})}(c) + \chi_{n-1}^{(i)}(c) \, \beta_{n-1}^{(i+2^{n-1})}
\notag \\
& \qquad \qquad \qquad \qquad \quad
{} + \chi_{n-1}^{(i)}(c) \, \chi_{n-1}^{(i+2^{n-1})}(c)
\label{eq:chi_minus}
\end{align}
for every $n \ge 1$ and $i \ge 1$, where
(a) follows from \eqref{def:chi};
(b) follows from \eqref{def:recursive_eps_lattice1};
(c) follows from the fact that $j \vee a = j \vee b$ implies $j \vee c = j \vee a$, as shown in the proof of \lemref{lem:sum_is_unity};
(d) follows from \lemref{lem:distributive}; and
(e) follows from the fact that (i) $k \wedge c = k \wedge b$ implies $k \vee a < k \vee b$ and (ii) $j \wedge c = j \wedge b$ and $j \vee a < j \vee b$ imply $j \vee c = j \vee a$, as shown in the proof of \lemref{lem:sum_is_unity}.

Finally, we observe that
\begin{align}
\theta_{n}^{(2i-1)}
& \overset{\mathclap{\text{(a)}}}{=}
1 - \Bigg( \beta_{n}^{(2i-1)}(a, b) + \sum_{c \in \mathcal{M}(a, b)} \chi_{n}^{(2i-1)}( c ) \Bigg)
\notag \\
& \overset{\mathclap{\text{(b)}}}{=} \,
1 - \Bigg( \beta_{n-1}^{(i)} \, \beta_{n-1}^{(i+2^{n-1})} + \sum_{c \in \mathcal{M}(a, b)} \chi_{n}^{(2i-1)}( c ) \Bigg)
\notag \\
& \overset{\mathclap{\text{(c)}}}{=} \,
1 - \Bigg( \beta_{n-1}^{(i)} \, \beta_{n-1}^{(i+2^{n-1})} + \sum_{c \in \mathcal{M}(a, b)} \Big( \beta_{n-1}^{(i)} \, \chi_{n-1}^{(i+2^{n-1})}(c)
\notag \\
& \qquad \qquad \quad
{} + \chi_{n-1}^{(i)}(c) \, \beta_{n-1}^{(i+2^{n-1})} + \chi_{n-1}^{(i)}(c) \, \chi_{n-1}^{(i+2^{n-1})}(c) \Big) \Bigg)
\notag \\
& \overset{\mathclap{\text{(d)}}}{=}
1 - \Bigg( \beta_{n-1}^{(i)} \, \beta_{n-1}^{(i+2^{n-1})} + \beta_{n-1}^{(i)} \sum_{c \in \mathcal{M}(a, b)} \chi_{n-1}^{(i+2^{n-1})}(c)
\notag \\
& \qquad \quad
{} + \beta_{n-1}^{(i+2^{n-1})} \sum_{c \in \mathcal{M}(a, b)} \chi_{n-1}^{(i)}(c) - C_{n}^{(i)}
\notag \\
& \qquad \qquad
{} + \bigg( \sum_{c \in \mathcal{M}(a, b)} \chi_{n-1}^{(i)}(c) \bigg) \, \bigg( \sum_{c \in \mathcal{M}(a, b)} \chi_{n-1}^{(i+2^{n-1})}(c) \bigg) \Bigg)
\notag \\
& =
1 + C_{n}^{(i)} - \bigg( \beta_{n-1}^{(i)} + \sum_{c \in \mathcal{M}(a, b)} \chi_{n-1}^{(i)}( c ) \bigg)
\notag \\
& \qquad \qquad \qquad \quad
{} \times \bigg( \beta_{n-1}^{(i+2^{n-1})} + \sum_{c \in \mathcal{M}(a, b)} \chi_{n-1}^{(i+2^{n-1})}( c ) \bigg)
\notag \\
& \overset{\mathclap{\text{(e)}}}{=}
1 + C_{n}^{(i)} - \Big( 1 - \theta_{n-1}^{(i)} \Big) \, \Big( 1 - \theta_{n-1}^{(i+2^{n-1})} \Big)
\notag \\
& =
\theta_{n-1}^{(i)} + \theta_{n-1}^{(i+2^{n-1})} - \theta_{n-1}^{(i)} \, \theta_{n-1}^{(i+2^{n-1})} + C_{n}^{(i)} ,
\end{align}
where
(a) follows from \lemref{lem:sum_is_unity};
(b) follows from \eqref{eq:beta_minus};
(c) follows from \eqref{eq:chi_minus};
(d) follows from \eqref{def:Cni}; and
(e) follows from \lemref{lem:sum_is_unity}.
This completes the proof of \lemref{lem:formulae}.
\hfill\IEEEQEDhere

\bibliographystyle{IEEEtran}
\bibliography{IEEEabrv,mybib}

\end{document}